\documentstyle[preprint,aps,12pt]{revtex}
\draft
\begin{document}
\bibliographystyle{unsrt}
\title{   Solution of the two-impurity Kondo model:
critical point, \\
Fermi-liquid phase, and crossover
}
\author{ Junwu Gan\thanks{Current address: Department of Physics,
	University of California, Berkeley, CA 94720.}
}
\address{
Department of Physics, The University of British Columbia,      \\
6224 Agricultural Road, Vancouver, B.C. Canada V6T 1Z1
}
\date{\today, \, \, \small UBCTP-94-008}

\maketitle

\begin{abstract}

An asymptotically exact solution is presented for the
two-impurity Kondo model for  a finite region
of the parameter space surrounding the critical point.
This region is  located in the  most
interesting intermediate regime
where  RKKY interaction is comparable to the
Kondo temperature.
 After several exact simplifications involving
reduction to one dimension and abelian bosonization,
the critical point is
explicitly identified, making clear its physical origin.
By using  controlled low energy projection,
an effective Hamiltonian is mapped out
for the finite region in the phase diagram around the critical point.
The completeness of the effective
Hamiltonian is rigorously proved from general symmetry considerations.
The effective Hamiltonian is solved exactly not only at
the critical point but also for the surrounding Fermi-liquid phase.
Analytic crossover functions
from  the critical to Fermi-liquid behavior are derived
for the specific heat and staggered susceptibility.
It is  shown that applying a uniform magnetic field has negligible
effect on the critical behavior.
A detailed comparison  is  made with the
numerical renormalization group
and conformal field theory results.
The excellent agreement is exploited to argue for
the  universality of the critical point, which in turn implies
universal behavior everywhere inside our  solution region.

\end{abstract}

\vskip .5in



\section{Introduction}

For a vast number of materials with strong electron correlation,
the low energy excitations involve both itinerant electrons
and well localized magnetic moments
residing periodically on the lattice sites. This is the case of
heavy fermion compounds~\cite{hfrev}, and to certain extent it is also the
case of high temperature superconducting cuprates~\cite{htcrev}.
In such systems,
two effects have crucial influence on the low energy properties
and they compete with each other.
They are the Kondo effect~\cite{kondo64}
 and RKKY interaction~\cite{rkky}, which
represent two different tendencies of the system to quench
the local moments with conduction electrons
or by themselves.
The simplest model capturing both effects is the
two-impurity Kondo model~\cite{jaya81,abra85,cole87}.
It is also believed that any possible new physics that
may occur in lattice due to the competition between the two
effects should be contained in the two-impurity problem~\cite{varma87}.

A simple way to see how the competition
arises in the two-impurity Kondo model
is to look at the problem from the  scaling point of view~\cite{ande70}.
The complexity of the problem
will be fully realized this way and the concrete
task we are facing will be defined.
For all  practical purposes, the bare Kondo coupling constant $J_{0}$
and the RKKY interaction $K_{0}$ are much smaller than the
Fermi energy $\epsilon_{F}$. Thus, if we form two dimensionless coupling
constants with the help of the
conduction electron density of states $\rho_{F}$($\sim 1/\epsilon_{F}$),
$\rho_{F} J_{0}$ and $\rho_{F} K_{0}$,
they are always in the weak coupling regime.
However, as we start to eliminate high energy conduction electron
states near the top and bottom of the conduction band,
both dimensionless coupling constants grow under renormalization
and they mutually renormalize each other.
Simple dimensional counting shows that $\rho_{F}J_{0}$ has dimension one
and is marginally relevant, while $\rho_{F} K_{0}$ has dimension zero and
is  relevant. For these two relevant interactions
we can define two energy scales, the Kondo temperature $T_{K}$
and RKKY temperature $T_{RKKY}$, such that they
correspond to the values of the decreasing conduction bandwidth
 at which the renormalized
dimensionless coupling constants $\rho_{F} J$ and
$\rho_{F} K$ reach the unity respectively.
In either $T_{K} \gg T_{RKKY}$ or $T_{K} \ll T_{RKKY}$ limit,
the problem is simple because we can perturb one of the two interactions.
The most difficult situation corresponds to $T_{K} \sim T_{RKKY}$.
This is also the situation  of
most practical interest.

Due to the broad interest in the competition between the Kondo effect
and RKKY interaction,
extensive investigations  have been carried out in the last
decade~\cite{jaya81,hirsh86,jones,rasul89,jones89,millis90,sakai90,yana91,sire93,affl94}.
As a result,
a convincing phase diagram has
emerged~\cite{jones,millis90,sakai90,yana91}, if not yet
universally accepted without reservation~\cite{fye94}.
This phase diagram is shown in Figure~\ref{phase_diag}.
The model exhibits Fermi-liquid behavior
everywhere except at a special point on the particle-hole
symmetric axis. At this point, the ratio between
the fully renormalized effective RKKY interaction and the Kondo temperature
is numerically estimated  to be $2.2$~\cite{jones}.
The effective RKKY interaction  is actually the RKKY temperature
whose meaning we have explained in the above.
But for some reason it has been confusingly called RKKY interaction.
Although
the precise numerical value of the ratio may depend on
the individual's convention of defining the coupling constants,
the important message is that this critical
point is located at  $T_{K} \sim T_{RKKY}$.

Although several asymptotically exact results have been available
 about the critical point in certain limits~\cite{jones,affl94},
the exact physical origin of this critical point had
not been unveiled until recently~\cite{gan94_3}.
By an explicit identification of
two local impurity  spin states whose level crossing being the origin
of the critical point,  we have rigorously shown
how the constraints set by the  discrete symmetries of the model
ensures the occurrence of a non-Fermi-liquid critical point.
We have also presented
an effective Hamiltonian for the finite
solution region as marked in Figure~\ref{phase_diag},
and listed  the  low temperature properties of the critical point.
In this paper, we present a detailed derivation  of
the effective Hamiltonian,
and  for the first time
 a full analytic solution for the  whole solution region
of Figure~\ref{phase_diag}.

Having admitted the existence of
the critical point, we can already present a framework
for the solution
inside that solution region  in
 Figure~\ref{phase_diag},
by  only invoking   general scaling ideas.
It is then the task of
 section~\ref{low-T-prop} to fill in concrete results.
Since Kondo effect always takes place in our solution region,
the basic energy scale must be
the Kondo temperature $T_{K}$,
which is much smaller than the Fermi energy.
On the scale of $T_{K}$, the system has already
lost its memory of microscopic details existed on the
energy scale of the Fermi energy.
The above mentioned $T_{RKKY}$ is of the same order as $T_{K}$,
and therefore does not constitute a new energy scale by itself.
However, the competition induces the second energy scale $T_{c}$.
Inside our solution region, $T_{c} \ll T_{K}$, and $T_{c}$
vanishes at the critical point. For any physical quantity,
its dependence on the bare parameters of the Hamiltonian should
be absorbed into these two energy scales. For instance, we
can write the specific heat
in the form $C(T) = f(T/T_{K}, T_{c}/T_{K})$,
where $f(x,y)$ is some  universal two-variable function.
The role of $T_{K}$ is simply to set an energy unit for the problem.
The second energy scale
$T_{c}$ determines the crossover from the non-Fermi-liquid
behavior governed by the critical point at $T \gg T_{c}$
to the Fermi-liquid behavior governed by the stable Fermi-liquid
fixed point at $T \ll T_{c}$.
If we recall how the appearance of the
Kondo temperature $T_{K}$($\ll \epsilon_{F}$)
 in the one-impurity Kondo problem
leads to the drastic enhancement of various physical quantities
including specific heat,
we can expect additional enhancement from the appearance
of $T_{c}$($\ll T_{K}$). The translation of this effect to the lattice
problem will be a new mechanism for the heavy electron mass.

Moving far away from the critical point
in the phase diagram, the accuracy of
our solution deteriorates. However, the
low energy exponents for all  physical
quantities   should  not change
since after all the system is still
governed by a Fermi-liquid fixed point,
as in the  solution region of  Figure~\ref{phase_diag}
near the critical point.
What need to be improved
are the constant prefactors.
Usually, a physical quantity calculated for a Fermi-liquid fixed
point is a sum of several contributions with the same exponent.
Among them, only a few are associated with the energy scale $T_{c}$,
while the others are associated with $T_{K}$.
What we calculate in this paper are those contributions associated
with the energy scale $T_{c}$. This is enough near the critical point
because they are enhanced inside our
solution region of Figure~\ref{phase_diag}.
As one moves away from the critical point, the other contributions
associated with the energy scale $T_{K}$ become
increasingly significant. Certainly, one can always fit the prefactors for
every physical quantity if well established numerical results
or experimental data are available. But it is the merit of the theory to
establish the relations between these prefactors in the same
spirit of Nozi\`{e}res's Fermi-liquid theory
of  Kondo effect~\cite{nozi74}.
A complete solution of the two-impurity Kondo model for
the whole parameter space is beyond the scope of this paper.

The layout of the paper is the following.
In section~\ref{red_1d}, we present the preliminary
transformations on the two-impurity Kondo model including
the reduction to one dimension and bosonization.
In section~\ref{mapping},
we identify the critical point and derive the effective Hamiltonian.
A rigorous proof of the completeness of the  effective
Hamiltonian is included.
In section~\ref{low-T-prop}, we solve the effective Hamiltonian
and calculate the low energy thermodynamics.
In section~\ref{compare}, we compare our results with those
derived from the numerical renormalization group and
conformal field theory approaches. The universality
of the critical point is strongly argued for.
We conclude our paper in section~\ref{conclusion} with a
 summary and some speculations on the lattice problem.
 To alleviate cross-referencing,
some frequently used parameters and symbols are
gathered in Table~\ref{notations}.

\section{Reduction to one dimension and Bosonization}  \label{red_1d}

In this section, we shall start from the most general two-impurity
Kondo model and perform various exact transformations to
reduce it to a simplified  form,
(\ref{h0_3})+(\ref{h1_3})+(\ref{h2_3}),
 suitable for identifying the
critical point and uncovering the underlying physics.

The general Hamiltonian for  the
anisotropic two-impurity Kondo model has the following
form,
\begin{eqnarray}
H &=&
\int d^{3}k \; \epsilon_{k}
\psi^{\dagger}_{\vec{k}} \psi_{\vec{k}}
+
\int \frac{ d^{3}k \, d^{3}k'}{ (2\pi)^{3} }
\sum_{\lambda=x,y,z} \frac{ J^{\lambda} }{2} \left[
e^{ \frac{i}{2} (\vec{k}-\vec{k}') \cdot\vec{R} }
\psi^{\dagger}_{\vec{k}} \sigma^{\lambda}
\psi_{\vec{k}'}  S^{\lambda}_{1}
+ e^{ - \frac{i}{2}(\vec{k}-\vec{k}') \cdot\vec{R} }
\psi^{\dagger}_{\vec{k}} \sigma^{\lambda}
\psi_{\vec{k}'}  S^{\lambda}_{2}
\right]
	\nonumber       \\
 & & \hspace{.1in}
 + \sum_{\lambda=x,y,z} K_{\lambda}
S^{\lambda}_{1} S^{\lambda}_{2}  ,
		\label{h1}
\end{eqnarray}
where $\psi^{\dagger}_{\vec{k}}
=( \psi^{\dagger}_{\vec{k}\uparrow},
\psi^{\dagger}_{\vec{k}\downarrow} )$,
$\sigma^{\lambda}$ with $\lambda=x,y,z$ are the Pauli matrices,
$\vec{S}_{1}$ and $\vec{S}_{2}$ are the two impurity spin 1/2
operators located at $\pm \vec{R}/2$.
It has been shown that the Hamiltonian~(\ref{h1}) can
be reduced to an equivalent  one-dimensional(1D)
problem~\cite{jones,affl94}.
 Introducing  1D fermionic operators,
\begin{equation}
\psi_{1,2}(k) = \frac{k}{\sqrt{2} }  \left[
\frac{1}{N_{e}(k)}
\int d^{2}\vec{\Omega} \,
\cos \left( \frac{\vec{k}\cdot\vec{R}}{2} \right)
\mp \frac{i}{N_{o}(k)}
\int d^{2}\vec{\Omega} \,
\sin \left( \frac{\vec{k}\cdot\vec{R}}{2} \right) \right]  \psi_{\vec{k}}  ,
	\label{fermion_1d}
\end{equation}
with  the notations $d^{3}k=k^{2}dk\, d^{2}\vec{\Omega}$ and
\begin{equation}
N_{e,o}(k) =
\sqrt{  1 \pm \frac{ \sin(kR) }{kR}  } ,
\end{equation}
we can completely rewrite Kondo interactions in terms of these new
operators,
\begin{eqnarray}
 H_{Kondo} &=&
\frac{v_{F}}{2}
\sum_{\lambda=x,y,x}
\int_{0}^{\infty} \frac{dk \; dk'}{(2\pi)^{2}}
\left\{
J_{+}^{\lambda}(k,k') \left[
\psi^{\dagger}_{1}(k) \sigma^{\lambda} \psi_{1}(k')
+  \psi^{\dagger}_{2}(k) \sigma^{\lambda} \psi_{2}(k')
\right] S_{+}^{\lambda}
	\right.         \nonumber       \\
 & &
 + J_{m}^{\lambda}(k,k') \left[
\psi^{\dagger}_{1}(k) \sigma^{\lambda} \psi_{1}(k')
 - \psi^{\dagger}_{2}(k) \sigma^{\lambda} \psi_{2}(k')
\right] S_{-}^{\lambda}
+  J_{-}^{\lambda}(k,k') \left[
\psi^{\dagger}_{1}(k) \sigma^{\lambda} \psi_{2}(k')
	\right.         \nonumber       \\
& &
\left. \left.
 + \psi^{\dagger}_{2}(k) \sigma^{\lambda} \psi_{1}(k')
\right] S_{+}^{\lambda}
-  J_{ir}^{\lambda}(k,k') \left[
\psi^{\dagger}_{1}(k) \sigma^{\lambda} \psi_{2}(k')
 - \psi^{\dagger}_{2}(k) \sigma^{\lambda} \psi_{1}(k')
\right] S_{-}^{\lambda}
\right\}  .
\end{eqnarray}
In the last expression,
we have  introduced the  short-hand notation,
\begin{equation}
S_{\pm}^{\lambda}=S_{1}^{\lambda} \pm S_{2}^{\lambda},
\hspace{.3in}   \lambda=x,y,z.          \label{spmxyz}
\end{equation}
The momentum dependent coupling constants are, for  $ \lambda=x,y,z $,
\begin{eqnarray}
	J_{+}^{\lambda}(k,k') &=&
\frac{J^{\lambda} k k'}{4\pi v_{F} } \left[
N_{e}(k) N_{e}(k') + N_{o}(k) N_{o}(k')
\right] ,               \\
	J_{-}^{\lambda}(k,k') &=&
\frac{J^{\lambda} k k'}{4\pi v_{F} } \left[
N_{e}(k) N_{e}(k') - N_{o}(k) N_{o}(k')
	\right],                \\
	J_{m}^{\lambda}(k,k') &=&
\frac{J^{\lambda} k k'}{4\pi v_{F} } \left[
N_{e}(k) N_{o}(k') + N_{o}(k) N_{e}(k')
	\right],                \\
	J_{ir}^{\lambda}(k,k') &=&
\frac{J^{\lambda} k k'}{4\pi v_{F} } \left[
N_{e}(k) N_{o}(k') - N_{o}(k) N_{e}(k')
\right] .
\end{eqnarray}
{}From the commutation relation $\{ \psi^{\dagger}_{\vec{k}\sigma},
\psi_{\vec{k}'\sigma'} \} = \delta^{3}(\vec{k}-\vec{k}')
\delta_{\sigma,\sigma'} $, we can verify
\begin{equation}
\{ \psi^{\dagger}_{i\sigma}(k), \psi_{i'\sigma'}(k') \}
= 2\pi \delta(k-k') \delta_{i,i'} \delta_{\sigma,\sigma'}.
\end{equation}
The free conduction electron Hamiltonian
can also be written in terms of these 1D fermions, plus completely
decoupled extra degrees of freedom,
\begin{equation}
\int d^{3}k \; \epsilon_{k}
\psi^{\dagger}_{\vec{k}} \psi_{\vec{k}}
= \int^{\infty}_{0} \frac{dk}{2\pi} \epsilon_{k}
\left[\psi_{1}^{\dagger}(k) \psi_{1}(k)
+ \psi_{2}^{\dagger}(k) \psi_{2}(k)      \right]    + \cdots .
\end{equation}
Thus, only 1D fermions defined in (\ref{fermion_1d})
are relevant to the two-impurity Kondo problem.

So far, the reduction has been exact.
In the next step, we  linearize
 the dispersion $\epsilon_{k} $ at $k=k_{F}$,
$\epsilon_{k} \simeq v_{F}(k-k_{F})$,
and expand the momentum dependent coupling constants
around $k=k_{F}$. We only need to retain the
leading terms of the expansion since
other terms contain some power of $k-k_{F}$, which  have
 high scaling dimension and are irrelevant at low energy.
{}From $J_{ir}^{\lambda}(k_{F},k_{F}) = 0$,
we see that $J_{ir}$ interaction only contains irrelevant terms.
Denoting $k-k_{F}$ by $k$ again and with implicit understanding
of an ultraviolet cutoff, we can
introduce Fourier transformations
\begin{equation}
\psi_{i}(x) = \int^{\infty}_{-\infty}
\frac{dk}{2\pi} e^{ikx} \psi_{i}(k) ,
\hspace{.3in} i=1,2.
\end{equation}
The fermion operators satisfy the standard commutation relation,
\begin{equation}
\{ \psi^{\dagger}_{i\sigma}(x), \psi_{i'\sigma'}(x') \} =
\delta(x-x') \delta_{i,i'} \delta_{\sigma,\sigma'} .
\end{equation}
After linearization,
the full two-impurity Kondo
Hamiltonian can be cast in the following form,
	\begin{eqnarray}                                                        H &=&
H_{0}+H_{1} ,        \label{hamil}      \\
H_{0} &=& -i \, v_{F} \sum_{i=1,2} \int^{\infty}_{-\infty}
dx \, \psi^{\dagger}_{i}(x) \partial_{x} \psi_{i}(x)
+ \sum_{\lambda=x,y,z} K_{\lambda} S^{\lambda}_{1} S^{\lambda}_{2}
	\nonumber       \\
 & + &  h_{u} \left[ S_{+}^{z}  + \frac{1}{2} \sum_{i=1,2}
 \int^{\infty}_{-\infty} dx
 \psi^{\dagger}_{i}(x) \sigma^{z} \psi_{i}(x) \right]
+ h_{s}  S_{-}^{z}     ,
	\label{h01}              \\
H_{1} &=&
 \frac{v_{F}}{2} \sum_{\lambda=x,y,z}
 \left\{  J_{+}^{\lambda}
\left[ \psi^{\dagger}_{1}(0) \sigma^{\lambda} \psi_{1}(0)
+  \psi^{\dagger}_{2}(0) \sigma^{\lambda} \psi_{2}(0) \right]
S_{+}^{\lambda}
	     \right.           \nonumber       \\
	     &  +  & \left.
 J_{m}^{\lambda}
\left[ \psi^{\dagger}_{1}(0) \sigma^{\lambda} \psi_{1}(0)
- \psi^{\dagger}_{2}(0) \sigma^{\lambda} \psi_{2}(0) \right]
S_{-}^{\lambda}
+  J_{-}^{\lambda}
\left[ \psi^{\dagger}_{1}(0) \sigma^{\lambda} \psi_{2}(0)
+  \psi^{\dagger}_{2}(0) \sigma^{\lambda} \psi_{1}(0) \right]
S_{+}^{\lambda}
  \right\} .
		\label{h1_0}
\end{eqnarray}
The coupling constants are, for $\lambda=x,y,z$,
\begin{equation}
J_{+}^{\lambda} = \pi J^{\lambda} \rho_{F}  ,
\hspace{.2in}
J_{-}^{\lambda} = \pi J^{\lambda} \rho_{F}
\frac{ \sin(k_{F}R ) }{ k_{F} R},
\hspace{.2in}
J_{m}^{\lambda} = \pi  J^{\lambda} \rho_{F}
\sqrt{1- \left( \frac{\sin k_{F}R }{k_{F}R} \right)^{2} } ,
	\label{coupl_const}
\end{equation}
with $\rho_{F}= k_{F}^{2}/(2\pi^{2} v_{F})$,
denoting the conduction electron density of states
per spin
at the Fermi energy.
Noting that we have  included both uniform
and staggered magnetic fields
$h_{u}$ and $h_{s}$
in (\ref{h01}), with Bohr magneton and gyromagnetic ratio set equal
to one.

At this stage, the 3D two-impurity Kondo model has been
successfully reduced to an equivalent 1D problem, up to some
terms irrelevant at low energy.
 However, we must remove
two accidental features
of (\ref{hamil}) resulting from linearization.
They are the particle-hole symmetry and a special relation
between
$J_{\pm}^{\lambda}(k_{F},k_{F})$ and $J_{m}^{\lambda}(k_{F},k_{F})$:
that the RKKY interaction generated
from them is always ferromagnetic~\cite{jones}.
These accidental features  will be spoiled by
 the generated corrections
from irrelevant terms neglected during the linearization.
That an irrelevant interaction can renormalize
the coupling constant of a relevant interaction
 is  a well known fact~\cite{ma75}.
An example can also be found
in  section~\ref{low-T-prop}, expression~(\ref{S2_1}),  where the
dimension 3/2 leading irrelevant
operator in the effective Hamiltonian~(\ref{heff}) induces
a correction to the  dimension 1/2 relevant operator.
Usually,  the accidental features
at the lowest order will not survive
if there is no hidden symmetry ensuring them.
 The accidental relation between
$J_{\pm}^{\lambda}(k_{F},k_{F})$ and $J_{m}^{\lambda}(k_{F},k_{F})$,
with $\lambda=x,y,z$,
is removed by treating these  coupling constants   as
{\em independent} parameters. This is also physically meaningful
since these  interactions are completely
independent and presumably play different roles
at low energy.
For a general conduction  band, particle-hole
symmetry breaking, although weak, is always present.
The  general particle-hole symmetry breaking term
that can be added to the 1D
Hamiltonian~(\ref{hamil}) has the following
form~\cite{jones,affl94},
\begin{equation}
  H_{2} = V \left[ \psi^{\dagger}_{1}(0) \psi_{2}(0)
+  \psi^{\dagger}_{2}(0)  \psi_{1}(0) \right] ,
	\label{ph-break}
\end{equation}
where
$V$ is the energy scale characterizing the strength of
particle-hole symmetry breaking. Adding the marginal
operator~(\ref{ph-break}) to the 1D Hamiltonian
after dropping irrelevant interactions in the linearization
may seem unusual, actually it is  the natural thing to do.
The reason is again the generation of (\ref{ph-break})
from irrelevant interactions in the absence of the
particle-hole symmetry.
Usually, all possible operators allowed by the symmetry
will be generated by irrelevant interactions,
and we only need to include relatively more relevant operators.
In this case,
the only non-irrelevant operator breaking particle-hole symmetry is
(\ref{ph-break}).  In section~\ref{mapping}
and appendix~\ref{app_proj},
we shall see a similar example where the marginal
operator~(\ref{ph-break}) generates
a relevant operator~(\ref{hphb}) when projecting to a subspace
relevant for the solution region of Figure~\ref{phase_diag}.

In the rest of
this paper,
we shall retain the rotational symmetry around  $z$-axis by
setting $K_{x}=K_{y}=K_{\perp}$
and $J_{i}^{x}=J_{i}^{y}=J_{i}^{\perp}$ for $i=m,\pm$.
Apart from the continuous U(1) rotational symmetry,
the Hamiltonian~(\ref{hamil}) possesses several discrete symmetries
which will be very useful for our analysis.
The transformation rules are, omitting unaffected operators,
\begin{eqnarray}
  {\rm Parity}  &:&  \hspace{.1in} \psi_{1} \leftrightarrow \psi_{2},
\hspace{.1in} S_{1}^{\lambda} \leftrightarrow  S_{2}^{\lambda}
\; \; {\rm for} \; \;  \lambda=x,y,z  ,
	\nonumber       \\
  {\rm Particle-hole} &:&  \hspace{.1in} \psi_{i\uparrow} \rightarrow
 \psi_{i\downarrow}^{\dagger}, \hspace{.1in}
\psi_{i\downarrow} \rightarrow
- \psi_{i\uparrow}^{\dagger}  ,
	  \label{symm}      \\
  \pi \; {\rm rotation \; around} \;
 x{\rm -axis} &:&  \hspace{.1in} \psi_{i\uparrow} \leftrightarrow
\psi_{i\downarrow}, \hspace{.1in} S_{i}^{y} \rightarrow -S_{i}^{y},
\; S_{i}^{z} \rightarrow -S_{i}^{z}    .
	\nonumber
\end{eqnarray}
The particle-hole symmetry exists when $V=0$.

   The next step is to
reduce the Hamiltonian~(\ref{hamil})+(\ref{ph-break})
to a simple form
   suitable for identifying the critical point.
   The reduction involves bosonizing the
 Hamiltonian which only contains 1D
 left-moving fermions~\cite{toul70,emery92,sire93}.
There are four species of fermions, so we need
to introduce four bose fields,
\begin{equation}
\psi_{i\sigma}(x) = \frac{P_{i\sigma}}{\sqrt{2\pi \alpha}}
e^{i \Phi_{i\sigma}(x)} , \hspace{.1in} i=1,2, \; \;
\sigma=\uparrow,\downarrow ,                   \label{bosoniz}
\end{equation}
where $\alpha$ is the lattice spacing and
\begin{equation}
\Phi_{i\sigma}(x)= \sqrt{\pi} \left[ \phi_{i\sigma}(x)-\int_{-\infty}^{x}
dx' \; \Pi_{i\sigma}(x') \right] .
	\label{def_Phi}
\end{equation}
The bose fields satisfy
 the standard commutation relation,
\begin{equation}
[\phi_{j\sigma}(x), \Pi_{j'\sigma'}(x')]=i \, \delta_{jj'}
\delta_{\sigma\sigma'} \delta(x-x')  .
\end{equation}
The phase factors
$P_{i\sigma}$ are introduced to take care of the anticommutation
relations between different species of fermions.
Our choices are,
\begin{eqnarray}
P_{1\uparrow} &=& P_{1\downarrow} =
e^{ i\pi \int^{\infty}_{-\infty}
dx \psi_{1\uparrow}^{\dagger}(x)
\psi_{1\uparrow}(x) } ,                    \\
 P_{2\uparrow} &=& P_{2\downarrow} =
e^{ i\pi \int^{\infty}_{-\infty} dx
 \left[ \sum_{\sigma}
\psi_{1\sigma}^{\dagger}(x)
\psi_{1\sigma}(x)  + \psi_{2\uparrow}^{\dagger}(x)
\psi_{2\uparrow}(x) \right] } .
\end{eqnarray}
By substituting (\ref{bosoniz}) into (\ref{hamil}) and  using the relation
$\psi_{i\sigma}^{\dagger}(x) \psi_{i\sigma}(x)
= \partial_{x} \Phi_{i\sigma}(x)/(2\pi)$, the two-impurity
Kondo Hamiltonian is expressed in terms of four bose fields
$\phi_{i\sigma}(x)$.
Then we make linear transformations to four new bose fields
corresponding to charge, spin, flavor and
spin-flavor degrees of freedom,
\begin{eqnarray}
\phi_{c} &=& (\phi_{1\uparrow}
+\phi_{1\downarrow}+\phi_{2\uparrow}
+\phi_{2\downarrow})/2  ,                   \nonumber   \\
\phi_{s} &=& (\phi_{1\uparrow}- \phi_{1\downarrow}
	+\phi_{2\uparrow}-\phi_{2\downarrow})/2 ,
					\nonumber     \\
\phi_{f} &=& (\phi_{1\uparrow}+ \phi_{1\downarrow}
	-\phi_{2\uparrow}-\phi_{2\downarrow})/2 ,        \\
\phi_{sf} &=& (\phi_{1\uparrow}- \phi_{1\downarrow}
	-\phi_{2\uparrow}+\phi_{2\downarrow})/2 .       \nonumber
  \end{eqnarray}
  The Hamiltonian now acquires the following form,
  \begin{eqnarray}
  H_{0} &=& \frac{v_{F}}{2} \sum_{\lambda=c,s,f,sf}
  \int^{\infty}_{-\infty}  dx \left\{ \Pi_{\lambda}^{2}(x)
  +[\partial_{x}\phi_{\lambda}(x)]^{2} \right\}
  + \sum_{\lambda=x,y,z} K_{\lambda} S^{\lambda}_{1} S^{\lambda}_{2}
	\nonumber               \\
  &  &
  + \, h_{u}\left[ S^{z}_{+}
+ \int^{\infty}_{-\infty} \frac{dx}{2\pi}
 \partial_{x} \Phi_{s}(x) \right]
 + h_{s} S^{z}_{-}  ,            \\
 H_{1} &=& \frac{v_{F}}{2} \left\{
   \frac{J^{z}_{+} }{\pi} \partial_{x} \Phi_{s}(0) S^{z}_{+}
 + \frac{J^{z}_{m} }{\pi} \partial_{x} \Phi_{sf}(0) S^{z}_{-}
 - i \frac{2J^{z}_{-} }{\pi\alpha} e^{i\pi \theta}
 \cos\Phi_{sf}(0) \sin\Phi_{f}(0) S^{z}_{+}
 + \frac{2J_{+}^{\perp}}{\pi\alpha}
			 \right.        \nonumber       \\
  &   \times &   \cos\Phi_{sf}(0)
    \left[ \cos\Phi_{s}(0) S^{x}_{+}
- \sin\Phi_{s}(0) S^{y}_{+} \right]
      - \frac{2J_{m}^{\perp}}{\pi\alpha} \sin\Phi_{sf}(0)
    \left[ \sin\Phi_{s}(0) S^{x}_{-}
+ \cos\Phi_{s}(0) S^{y}_{-} \right]
					\nonumber  \\
  & &   \left.
  - \frac{2J_{-}^{\perp}}{\pi\alpha} e^{i\pi\theta}   \sin\Phi_{f}(0)
    \left[ \sin\Phi_{s}(0) S^{x}_{+}
+ \cos\Phi_{s}(0) S^{y}_{+} \right]
					\right\} ,
   \label{h1_1}     \\
  H_{2} &=&  -i \frac{2V}{\pi\alpha} e^{i\pi\theta}
	\sin\Phi_{sf}(0)  \cos\Phi_{f}(0)  ,
  \end{eqnarray}
 where the phase factor is
\begin{equation}
\theta=\int^{\infty}_{-\infty} dx [\psi_{1\downarrow}^{\dagger}(x)
\psi_{1\downarrow}(x) + \psi_{2\uparrow}^{\dagger}(x)
\psi_{2\uparrow}(x) ] = \frac{1}{2\pi} \int^{\infty}_{-\infty}
dx [\partial_{x}\Phi_{c}(x)
- \partial_{x}\Phi_{sf}(x) ] .
\end{equation}
The charge bose field $\phi_{c}(x)$ is decoupled
from  the interaction~(\ref{h1_1}).
It will be omitted
from now on, and so will be  the $\partial_{x} \Phi_{c} $ term
inside the integral of  the phase $\theta$.
The  $\cos\Phi_{s}(0)$ and $\sin\Phi_{s}(0)$ factors
in (\ref{h1_1}) can
be eliminated by  rotating the impurity spins around  $z$-axis
by an angle $\Phi_{s}(0)$,
\begin{eqnarray}
  &  & H \rightarrow \hat{U} H \hat{U}^{-1} ,  \hspace{.4in}
   {\rm with} \; \;
    \hat{U}=e^{-i S_{+}^{z} \Phi_{s}(0) },              \\
 & & \hat{U} H_{0} \hat{U}^{-1} =
  \frac{v_{F}}{2} \sum_{\lambda=s,f,sf}
  \int^{\infty}_{-\infty}  dx \left\{ \Pi_{\lambda}^{2}(x)
  +[\partial_{x}\phi_{\lambda}(x)]^{2} \right\}
  + \sum_{\lambda=x,y,z} K_{\lambda} S^{\lambda}_{1} S^{\lambda}_{2}
  + h_{s} S^{z}_{-}
		\nonumber               \\
  & & \; \; \; \; \;  +  h_{u} \int^{\infty}_{-\infty} \frac{dx}{2\pi}
 \partial_{x} \Phi_{s}(x)
   - v_{F} \partial_{x}\Phi_{s}(0) S^{z}_{+}
   + \frac{v_{F}}{\alpha} ( S^{z}_{+} )^{2}  ,
	 \label{UH0U}       \\
 & & \hat{U} H_{1} \hat{U}^{-1} =
 \frac{v_{F}}{2} \left\{
   \frac{J^{z}_{+} }{\pi} \partial_{x} \Phi_{s}(0) S^{z}_{+}
 + \frac{J^{z}_{m} }{\pi} \partial_{x} \Phi_{sf}(0) S^{z}_{-}
 - i \frac{2J^{z}_{-} }{\pi\alpha} e^{i\pi \theta}
 \cos\Phi_{sf}(0) \sin\Phi_{f}(0) S^{z}_{+}
			 \right.        \nonumber       \\
 & & \; \; \; \;  - \left. \frac{2 J^{z}_{+} }{\pi\alpha}
(S^{z}_{+})^{2}
  + \frac{2}{\pi\alpha} \left[
  J_{+}^{\perp} \cos\Phi_{sf}(0) S^{x}_{+}
  - J_{m}^{\perp} \sin\Phi_{sf}(0) S^{y}_{-}
  - J_{-}^{\perp} e^{i\pi\theta}   \sin\Phi_{f}(0)
  S^{y}_{+} \right]
			   \right\} .     \label{H1_2}
 \end{eqnarray}
 The particle-hole symmetry breaking term
 $H_{2}$ is not affected by the above rotation.
 We note that  both $J^{z}_{+}$ and $K_{z}$
 acquired corrections under the rotation,
and $h_{u} S^{z}_{+}$  is canceled out
in (\ref{UH0U}).
  The interactions in (\ref{H1_2}) only contain
  $\partial_{x}\Phi_{s}(0)$.   Therefore
the bose field $\phi_{s}(x)$ can be integrated out
analytically upon our wish.

The bosonized Hamiltonian~(\ref{UH0U})+(\ref{H1_2})
can be re-fermionized by
introducing three  species of fermions,
\begin{eqnarray}
 & & \psi_{sf}(x)=\frac{1}{\sqrt{2\pi\alpha}}
 e^{ i\Phi_{sf}(x) } ,           \\
 & & \psi_{f}(x) e^{ i\pi \int^{\infty}_{-\infty}
 dx \psi_{sf}^{\dagger}(x)
\psi_{sf}(x) } = \frac{1}{\sqrt{2\pi\alpha}}
 e^{ i\Phi_{f}(x) } ,          \\
 & & \psi_{s}^{\dagger}(x)\psi_{s}(x) =
 \frac{1}{2\pi} \, \partial_{x}\Phi_{s}(x) .
 \end{eqnarray}
Again, a phase factor is included in the definition of the
fermion operator $\psi_{f}(x)$ to take care of the
anticommutation relations between three different species
of fermions.  Because the interactions in (\ref{H1_2})
contain only
$\partial_{x}\Phi_{s}(0)$($=2 \pi \psi_{s}^{\dagger}(0)\psi_{s}(0)$),
so we do not
need to specify the phase for the fermion operator $\psi_{s}(0)$.
The complete Hamiltonian can be rewritten as
\begin{eqnarray}
H_{0} &=& -iv_{F} \sum_{i=s,f,sf} \int^{\infty}_{-\infty}
dx \psi_{i}^{\dagger}(x)
\partial_{x} \psi_{i}(x)
+ \widetilde{K}_{z} S^{z}_{1} S^{z}_{2}
+ K_{\perp} \sum_{\lambda=x,y} S^{\lambda}_{1} S^{\lambda}_{2}
				\nonumber       \\
 & & + h_{u} \int^{\infty}_{-\infty}
 dx \psi_{s}^{\dagger}(x) \psi_{s}(x)
+h_{s} S_{-}^{z}   ,
	\label{h0_3}         \\
 H_{1} &=&
 \frac{v_{F}}{2} \left\{ \widetilde{J}_{+}^{z} [\psi_{s}^{\dagger}(0)
\psi_{s}(0) - \psi_{s}(0)\psi_{s}^{\dagger}(0) ] S_{+}^{z}
+ J_{m}^{z} [ \psi_{sf}^{\dagger}(0)
\psi_{sf}(0) - \psi_{sf}(0) \psi_{sf}^{\dagger}(0) ] S_{-}^{z}
	\right.         \nonumber               \\
 & & \left. + J_{-}^{z} [ \psi_{sf}(0) + \psi_{sf}^{\dagger}(0) ]
[ \psi_{f}(0) - \psi_{f}^{\dagger}(0) ] S_{+}^{z}
\right\}
 +  \frac{v_{F}}{\sqrt{2\pi\alpha}}
\left\{ J_{+}^{\perp} [ \psi_{sf}(0) + \psi_{sf}^{\dagger}(0) ] S_{+}^{x}
	\right.         \nonumber       \\
& & \left. +i J_{m}^{\perp}
[ \psi_{sf}(0) - \psi_{sf}^{\dagger}(0) ] S_{-}^{y}
+i J_{-}^{\perp}
[ \psi_{f}(0) - \psi_{f}^{\dagger}(0) ] S_{+}^{y} \right\} ,
	     \label{h1_3}
\end{eqnarray}
where
\begin{equation}
\widetilde{J}_{+}^{z}  = J_{+}^{z}-2\pi ,             \hspace{.5in}
\widetilde{K}_{z} = K_{z}
 -  \frac{ 2 v_{F} }{ \pi\alpha } \left( J_{+}^{z} - \pi \right) .
\label{tildeK}
\end{equation}
The particle-hole symmetry breaking term becomes
\begin{equation}
H_{2} = V \left[ \psi_{sf}(0)-\psi_{sf}^{\dagger}(0) \right]
 \left[ \psi_{f}(0)+\psi_{f}^{\dagger}(0) \right] .     \label{h2_3}
\end{equation}

How do these new fermion operators transform under the discrete
symmetries of (\ref{symm})?
We can keep track of the transformation
rules during the bosonization and
subsequent fermionization to derive, omitting unaffected operators,
\begin{eqnarray}
  {\rm Parity}  &:&  \hspace{.1in}
  \psi_{sf} \leftrightarrow \psi_{sf}^{\dagger},
\hspace{.1in}
 \psi_{f} \leftrightarrow - \psi_{f}^{\dagger},
 \hspace{.1in}
 S_{1}^{\lambda} \leftrightarrow S_{2}^{\lambda}
\; \; {\rm for} \; \; \lambda=x,y,z ,
	\nonumber       \\
  {\rm Particle-hole} &:&  \hspace{.1in}
 \psi_{f} \leftrightarrow - \psi_{f}^{\dagger} ,
	   \label{symm1}     \\
  \pi \; {\rm rotation,} \;
 x{\rm -axis} &:&  \hspace{.1in}
  \psi_{sf} \leftrightarrow \psi_{sf}^{\dagger},
\hspace{.1in}
  \psi_{s} \leftrightarrow \psi_{s}^{\dagger},
\hspace{.1in}
 \psi_{f} \leftrightarrow - \psi_{f},
 \hspace{.1in}
 S_{i}^{y} \rightarrow - S_{i}^{y} ,
 \hspace{.1in}
 S_{i}^{z} \rightarrow - S_{i}^{z} .
	\nonumber
\end{eqnarray}
Alternatively, one can directly
 verify them from (\ref{h0_3}), (\ref{h1_3}) and (\ref{h2_3}).

\section{Effective Hamiltonian near the critical point} \label{mapping}

In this section,
we shall identify the critical point from
(\ref{h0_3})+(\ref{h1_3}) and derive an effective Hamiltonian
for the finite region of
 the parameter space surrounding the critical point.
Our identification of the critical point will make clear its physical
origin.

To search for the critical point,
we need only consider the particle-hole symmetric case, $V=0$.
 At first glance
the Hamiltonian (\ref{h0_3})+(\ref{h1_3}) still looks
too complicated to provide any intuition.
On the other hand,
from the conformal field theory results it is known that
the critical point  exists in a restricted Hamiltonian with
$J^{z}_{-}=J_{-}^{\perp}=0$, and around the critical point
$J^{z}_{-}$, $J_{-}^{\perp}$ interactions are irrelevant.
Thus, our task is greatly reduced by
searching the critical point in this restricted Hamiltonian.
An important step is to verify the irrelevance
of the $J^{z}_{-}$, $J_{-}^{\perp}$
interactions after we find the critical point.

We  have  noted before hat
only the product
$\psi_{s}^{\dagger}(0) \psi_{s}(0) $ appears in (\ref{h1_3}).
This  is because only
  $\partial_{x}\Phi_{s}(0)$
  appears in (\ref{H1_2}) and there isn't anything containing
  $\cos \Phi_{s}(0)$ or $\sin \Phi_{s}(0)$.
Thus, the $\widetilde{J}^{z}_{+}$ term containing
the bose field $\phi_{s}(x)$ can be integrated out analytically.
       Although $\partial_{x}\Phi_{s}(0)$ couples to an operator
       $S_{+}^{z}$,
       the integration can  be done formally in the
       path integral formalism, yielding
the following two terms to the action,
\begin{equation}
- \frac{ v_{F}(\widetilde{J}_{+}^{z})^{2} } {4\pi^{2}\alpha}
       \int^{\beta}_{0} d\tau
       [ S_{+}^{z}(\tau)]^{2} +
       \frac{ ( \widetilde{J}_{+}^{z} )^{2} }{ 8\pi } \sum_{n}
      |\nu_{n}| S_{+}^{z}(-\nu_{n})
      S_{+}^{z}(\nu_{n}) ,  \;\;
       {\rm with} \;\; \nu_{n}=2n\pi/\beta .
\end{equation}
The first term is a correction
to the RKKY interaction, so it is absorbed into
$\widetilde{K}_{z}$.
The second term has higher dimension and is expected to be irrelevant.
The point we want to make here
is that the $\widetilde{J}^{z}_{+}$ interaction does not
affect the critical point and can be ignored during the search
for the critical point.
Thus, we see that
when $J^{z}_{-}=J_{-}^{\perp}=0$
the remaining Kondo interactions only involve
three local spin operators, $S_{+}^{x}$,
$S_{-}^{y}$, $S_{-}^{z}$, which only act on three(out of four)
 local impurity spin states.
The impurity spin state
$ ( |\uparrow\uparrow> - \, |\downarrow\downarrow> )/\sqrt{2} $
decouples from the Kondo interactions
when $J^{z}_{-}=J_{-}^{\perp}=0$.
Together with
 the RKKY interactions,
we derive an energy level scheme for
 the impurity spin states in  Figure~\ref{loc_lev}.
 The critical point corresponds to the
 special case when the two lowest levels become degenerate.
Specifically,
 the $- \widetilde{K}_{z}(S_{-}^{z})^{2}/2 $ term in (\ref{h0_3})
 raises the energy of the
 states
 $|\uparrow\downarrow>$
 and  $|\downarrow\uparrow>$
 by an amount
 $- \widetilde{K}_{z}/2 $(assuming $ - \widetilde{K}_{z} > 0 $)
with respect to the other two states
  $|\uparrow\uparrow> \pm  |\downarrow\downarrow>$.
 These two  states,
 $|\uparrow\downarrow>$
 and  $|\downarrow\uparrow>$,
 are further split symmetrically
 by the transverse part of the RKKY interaction,
 $ K_{\perp} ( S_{1}^{+}  S_{2}^{-}
 + S_{1}^{-}  S_{2}^{+} )/2 $.
 When  $-\widetilde{K}_{z}=K_{\perp}$, the two levels,
 $ ( |\uparrow\downarrow> - \, |\downarrow\uparrow> )/\sqrt{2} $
 and
 $ ( |\uparrow\uparrow> + \, |\downarrow\downarrow> )/\sqrt{2} $,
 become degenerate, forming a doublet.
 Because there is almost no Kondo interaction
 in the  state
 $ ( |\uparrow\uparrow> - \, |\downarrow\downarrow> )/\sqrt{2} $
when $J^{z}_{-}=J_{-}^{\perp}=0$,
 the superficial degeneracy between this state and the  doublet
 is lifted by the
 Kondo interactions in the doublet
 which lower the energy of
 the doublet by a finite amount $T_{K}$, equal to
  the ground state energy gain at the critical point.
Turning on $J^{z}_{-}$ and $J_{-}^{\perp}$ will not change
the energy level scheme as long as
$v_{F}J^{z}_{-}, \; v_{F}J_{-}^{\perp} < T_{K}$.

  Now, we have identified the critical point.
 To describe the low energy physics, it is sufficient to project
 the full interacting
Hamiltonian~(\ref{h0_3})+(\ref{h1_3})+(\ref{h2_3})
 onto the
 lowest energy doublet.
 For $\widetilde{J}_{+}^{z}$,  $J_{+}^{\perp}$, $J_{m}^{z}$,
 $J_{-}^{z}$, $J_{-}^{\perp}$, $V$
 all but $J_{m}^{\perp}$ much smaller than one,
 the projection can be done accurately.
 Let $\hat{Q}$ be the projection operator onto the doublet,
 the projected effective Hamiltonian to the second order
 is
 \begin{equation}
 H_{eff} = \hat{Q} H \hat{Q}
 + \hat{Q} H ( 1- \hat{Q} )
 \frac{1}{  E_{0} - \hat{Q} H \hat{Q}
 - (1-\hat{Q}) H (1-\hat{Q})   }
 ( 1- \hat{Q} ) H \hat{Q}   ,         \label{proj2}
 \end{equation}
 where $E_{0}$ is ground state energy.
 The doublet can be described by local  fermion operators
 $d$ and $d^{\dagger}$ such that
 $ ( |\uparrow\uparrow> + \, |\downarrow\downarrow> )/\sqrt{2} $
 and
  $ ( |\uparrow\downarrow> - \, |\downarrow\uparrow> )/\sqrt{2} $
  correspond to $d^{\dagger}d=0$ and $d^{\dagger}d=1$
  states respectively.
 From Figure~\ref{loc_spin},   it is not difficult to verify
 $ \hat{Q} S_{-}^{y} \hat{Q} = i(d-d^{\dagger})$,
 $ \hat{Q} S_{+}^{y} \hat{Q} = \hat{Q} S_{+}^{x} \hat{Q} =
  \hat{Q} S_{\pm}^{z} \hat{Q} = 0 $.
 These relations are used for the first order projection.
 In the second order projection,  nonvanishing
 terms may contain
 $\hat{Q} (S_{-}^{z})^{2} \hat{Q} = d^{\dagger}d$,
   $\hat{Q} (S_{+}^{x})^{2} \hat{Q}=
    \hat{Q} (S_{+}^{z})^{2} \hat{Q}
   =d d^{\dagger}$,
    $\hat{Q} S_{+}^{x} S_{-}^{z}   \hat{Q} = d$
     and
    $\hat{Q} S_{-}^{z} S_{+}^{x}   \hat{Q} = d^{\dagger} $.
 Since the extended fermions commute with the impurity spin operators,
 we need to
 to install anticommutation relations between
 the local fermion operators $d$, $d^{\dagger}$ and the extended
 fermion operators $\psi_{\lambda}(x)$ with $\lambda=s,f,sf$.
 This is accomplished by a simple transformation,
 \begin{equation}
 \psi_{\lambda}(x) = \widetilde{ \psi}_{\lambda}(x)  \,
 e^{i\pi d^{\dagger}d } ,
\hspace{.3in}  \lambda=s,f,sf.            \label{ch_sta}
 \end{equation}
 The commutation relations between $d$ and $\psi_{\lambda}(x)$
 are converted to anticommutation relations between $d$
 and $\widetilde{\psi}_{\lambda}(x)$.
 The effective Hamiltonian will be represented in terms
 of $d$ and $\widetilde{\psi}_{\lambda}(x)$.
 But we shall omit the tilde signs on $\psi_{\lambda}(x)$ in the following.
With the help of the above mentioned results,
it is straightforward to evaluate (\ref{proj2}). The details
are captured in appendix~\ref{app_proj}.
The results are
\begin{eqnarray}
H_{eff} &=& H_{fp}+H_{pert} + H_{phb} ,       \label{heff}    \\
H_{fp} &=& -iv_{F} \sum_{i=f,sf} \int^{\infty}_{-\infty}
dx \psi_{i}^{\dagger}(x)
\partial_{x} \psi_{i}(x)
 + v_{F} g_{0}
 \left[ \psi_{sf}(0)-\psi_{sf}^{\dagger}(0) \right]  (d+d^{\dagger})
	\nonumber       \\
 & & + \alpha_{u} h_{u}
 \left[ \psi_{sf}(0) +\psi_{sf}^{\dagger}(0) \right]
\left[  \psi_{f}(0) - \psi_{f}^{\dagger}(0)  \right]
+\alpha_{s}  h_{s}
\left[ \psi_{sf}(0)+\psi_{sf}^{\dagger}(0) \right]  (d-d^{\dagger})  ,
	\label{HFP}          \\
H_{pert} &=&
	- \left( \frac{K_{\perp}+K_{z} }{2} - K_{c} \right)
d^{\dagger} d
- i  v_{F} g_{1} \,  (d-d^{\dagger})
\partial_{x} \left[ \psi_{sf}(x)-\psi_{sf}^{\dagger}(x) \right]_{x=0} ,
	\label{HPERT}   \\
 H_{phb} &=&
 \widetilde{V} [\psi_{f}(0)+\psi_{f}^{\dagger}(0)] (d - d^{\dagger}) ,
		\label{hphb}
\end{eqnarray}
where $K_{c}$ is the critical value of $(K_{z}+K_{\perp})/2$.
We have separated the Hamiltonian into the fixed point part,
 perturbation part, and particle-hole symmetry
breaking part.  We note that  (\ref{HFP})
in the absence of the external magnetic fields
$h_{u}$ and $h_{s}$ has the same form as that
of the two-channel Kondo model~\cite{emery92}.
To the second order,
the coefficients  in $H_{eff}$ are given by
\begin{eqnarray}
g_{0} &=& - \frac{ J_{m}^{\perp} }{ \sqrt{2\pi\alpha} }
\left[ 1+  \frac{ v_{F}J_{+}^{\perp} J_{m}^{z} }{
 4\pi\alpha (K_{\perp}+T_{K}) J_{m}^{\perp} } \right] ,
			\label{def_g0}       \\
g_{1} &=&  \frac{ v_{F}^{2}J_{+}^{\perp}
J_{m}^{z} }{ 8 (K_{\perp}+T_{K})^{2} (2\pi\alpha)^{3/2} },
			\label{def_g1}      \\
\widetilde{V} &=&  \frac{ v_{F}^{2} J_{+}^{\perp} J_{m}^{z} V
 }{ 2 (K_{\perp}+T_{K})^{2} (2\pi\alpha)^{5/2}  } ,
			\label{def_tilde_v}             \\
\alpha_{u} &=&  \frac{ v_{F} J_{-}^{z} \widetilde{J}_{+}^{z}
}{ 8 \pi T_{K} },         \label{def_alp_u}      \\
\alpha_{s} &=& -
\frac{ v_{F}J_{+}^{\perp} }{ (K_{\perp}+T_{K})\sqrt{2\pi\alpha} } .
			\label{def_alp_s}
\end{eqnarray}
The projection induces corrections to the RKKY interactions
 so the critical value  $K_{c}$ is not exactly
$v_{F} (J_{+}^{z}-\pi) /(\pi\alpha)$, as
determined by the condition
$-\widetilde{K}_{z}=K_{\perp}$.
The energy  $T_{K}$ in the above expressions can be identified
as the ground state energy gain in (\ref{HFP})
from the hybridization term $v_{F} g_{0}$.
{}From the study of the two-channel Kondo problem, we know
$T_{K} \sim v_{F} g_{0}^{2}$.

 The spin degrees of freedom are
completely decoupled and their Hamiltonian is,
parallel to (\ref{heff}),
\begin{equation}
H_{s} = -iv_{F} \int^{\infty}_{-\infty}
dx \psi_{s}^{\dagger}(x)
\partial_{x} \psi_{s}(x)
+ h_{u} \int^{\infty}_{-\infty}
 dx \psi_{s}^{\dagger}(x) \psi_{s}(x)
+ \frac{ v_{F}  (\widetilde{J}_{+}^{z})^{2}  h_{u} }{4\pi T_{K} }
\psi^{\dagger}_{s}(0) \psi_{s}(0) .
\end{equation}
Since this piece of Hamiltonian does not contain
any interesting physics, we shall not discuss it hereafter.
For the particle-hole symmetry breaking and staggered field
coupling terms, apart from  the relevant operators
we have included in (\ref{HFP}) and (\ref{hphb}),
there are also marginal operators
such as (\ref{h2_3}). They will be
discussed below and in appendix~\ref{more_chis}.

The effective Hamiltonian~(\ref{heff}) is the central result
of this paper.
The rest of this section is to prove the completeness of the
(\ref{heff}) from general symmetry considerations.
 For a certain region of the parameter space, the projection
 is controllable in a sense that high order corrections to the
 coefficients of $H_{eff}$ are too small to alter its critical
 behavior.
 However, the projection is not done exactly.
One may ask how do we know
 that there are no other operators which could arise from
  high orders of the  projection and spoil the critical behavior?
 Fortunately, it turns out that
 all other operators up to dimension 3/2 inclusive
 can be eliminated by
 the three discrete symmetries of (\ref{symm1}).
 To show this, we first determine how
 the operators $d$ and $d^{\dagger}$ transform under
 parity and the rotation of an angle $\pi$ around $x$-axis.
 The particle-hole transformation does not involve
 the impurity spins, so will not affect $d$ and $d^{\dagger}$.
 Since the two states of the doublet have different parity
 and $d^{\dagger}$, $d$ connect them,  we conclude that under parity:
 $d^{\dagger} \rightarrow -d^{\dagger}$,
 $d \rightarrow - d$.   This could also be seen
 from
 $\hat{Q} S_{-}^{z} S_{+}^{x}   \hat{Q} = d^{\dagger} $
     and
 $\hat{Q} S_{+}^{x} S_{-}^{z}   \hat{Q} = d $. From them,
 we can also see that under the $\pi$ rotation around $x$-axis,
      $d^{\dagger} \rightarrow -d^{\dagger}$, $ d \rightarrow - d$.
Combining these results with (\ref{symm1}),
we derive all ``elementary operators'' in the
projected Hilbert space
comprised of the local doublet
and the extended fermions $\psi_{\lambda}(x)$ with $\lambda=s,f,sf$.
They are listed in Table~\ref{blocks}.

Some explanations are necessary at this point.
First,  it is well known that 1D extended fermion operators
 have scaling dimension 1/2. This can be easily seen from the
 free fermion action $S(\psi,\psi^{\dagger})= \int^{\beta}_{0} d\tau
 \int^{\infty}_{-\infty} dx \psi^{\dagger}(x,\tau) (\partial_{\tau}
 -i v_{F} \partial_{x} ) \psi(x,\tau)$.
 Secondly, we
have noted before that $\psi_{s}$ and $\psi_{s}^{\dagger}$ can only
appear in the product  $\psi_{s}^{\dagger}$ $\psi_{s}$.
Therefore, the spin degrees of freedom do not bring in
the dimension 1/2  operators $\psi_{s}$ and $\psi_{s}^{\dagger}$
as additional building blocks in Table~\ref{blocks}.
Thirdly,  usual local fermion operators have scaling dimension zero.
As can be seen from the
free fermion action $ S(d,d^{\dagger})= \int^{\beta}_{0}
d\tau d^{\dagger}(\tau) \partial_{\tau} d(\tau) $,
we need not change $d$ and $d^{\dagger}$ under a rescaling
of the imaginary time $\tau$.
This would imply that both combinations $d \pm d^{\dagger}$
have scaling dimension zero. However,
the dimension of the operator
combination $d+d^{\dagger}$ is
raised to 1/2 by the hybridization term in (\ref{HFP})
with coefficient $v_{F} g_{0}$.
This follows immediately from the requirement of
preserving scale invariance of the hybridization term
under a rescaling of $x$ and $\tau$.
Lastly,
because  both $d$ and $d^{\dagger}$ are odd under parity
and $\pi$ rotation around $x$-axis,
only  $\psi_{sf}(0)-\psi_{sf}^{\dagger}(0)$
could hybridize with them
to give rise to a term in the effective Hamiltonian
even under all discrete
symmetries,  as can be seen from Table~\ref{blocks}.
Out of two linear independent combinations from $d$ and $d^{\dagger}$,
$\psi_{sf}(0)-\psi_{sf}^{\dagger}(0)$  could
only hybridize with one. Requiring it to be hermitian,
the hybridizing combination could be either
$d+d^{\dagger}$  or $d-d^{\dagger}$.
Thus, once $[\psi_{sf}(0)-\psi_{sf}^{\dagger}(0)](d+d^{\dagger})$
is generated in (\ref{HFP}),
$(d-d^{\dagger})[\psi_{sf}(0)-\psi_{sf}^{\dagger}(0)]$
is forbidden.   This guarantees that the dimension of
$d-d^{\dagger}$ will remain to be zero, like usual
local operators.

To construct
operators
in the projected Hilbert space, we only need to multiply together
the building blocks
of Table~\ref{blocks} and keep products of even number of fermionic
operators.
We list all dimension 1/2 operators in Table~\ref{dim12},
all dimension 1 operators in Table~\ref{dim1}, and
all dimension 3/2 operators in Table~\ref{dim32}.

Let us first consider the
 particle-hole
symmetric case.
Any  operator
that could appear in the effective Hamiltonian
must be even under all three
discrete symmetry operations~(\ref{symm1}).
We can explicitly verify that
all allowed operators up to dimension 3/2  that could
appear in the effective Hamiltonian
have been included in
(\ref{HFP}) and (\ref{HPERT}).
In order to couple to the uniform magnetic field,
an operator  has to be
even under parity and
 particle-hole transformations
 but odd under $\pi$ rotation around $x$-axis.
{}From Tables~\ref{dim12} and \ref{dim1}, we also
 verify that the only allowed operator up to dimension 1 inclusive
 is the one appearing in (\ref{HFP}).
 As to the operators that could couple to the staggered magnetic field,
 they must be even under particle-hole transformation but
 odd under parity transformation and $\pi$ rotation around $x$-axis.
Apart from the dimension 1/2 operator that couples
to the staggered field in (\ref{HFP}),  two more
dimension 1 operators are allowed by the symmetries.
They are the third and fourth operators in Table~\ref{dim1}.
Thus, including marginal operators we could
have the following
additional staggered field coupling terms added to (\ref{heff}),
\begin{equation}
H'_{stag} = \alpha'_{s} h_{s}
\left[ \psi_{sf}(0)+\psi_{sf}^{\dagger}(0) \right]
\left[ \psi_{sf}(0)-\psi_{sf}^{\dagger}(0) \right]
+ i \alpha''_{s} h_{s}
\left[ \psi_{sf}(0)+\psi_{sf}^{\dagger}(0) \right]
(d+d^{\dagger}) ,               \label{HSTAG_MORE}
\end{equation}
where $\alpha'_{s}$ and $\alpha''_{s}$ are two
dimensionless parameters
depending on the original coupling constants of (\ref{hamil}).
 Nevertheless,
the  contributions to the staggered
 susceptibility   from  (\ref{HSTAG_MORE}) are negligible
 as we shall see in appendix~\ref{more_chis}.

 A subtle point  arises here. We have used the argument,
that $\psi_{sf}(0)-\psi_{sf}^{\dagger}(0)$ could hybridize
with either one of $d \pm d^{\dagger}$ but not both,
to rule out possible hybridization between
$\psi_{sf}(0)-\psi_{sf}^{\dagger}(0)$
and $d - d^{\dagger}$ in (\ref{HFP}).
Why couldn't we make the same argument to eliminate
$ [\psi_{sf}(0)+\psi_{sf}^{\dagger}(0)]
(d+d^{\dagger}) $ in (\ref{HSTAG_MORE}),
since we already have
$ [\psi_{sf}(0)+\psi_{sf}^{\dagger}(0)]
(d-d^{\dagger}) $ in (\ref{HFP}).
The reason is the following.
For an arbitrary hybridization, we can rewrite it
in the following way
\[ \left[ \psi_{sf}(0)-\psi_{sf}^{\dagger}(0) \right]
 \left[ \alpha  (d-d^{\dagger})
+ i \beta (d+d^{\dagger})  \right]
= \sqrt{ \alpha^{2}+\beta^{2} }
\left[ \psi_{sf}(0)-\psi_{sf}^{\dagger}(0) \right]
\left( d \, e^{i\varphi}  - d^{\dagger} \, e^{-i\varphi} \right) ,   \]
where $\alpha, \; \beta$ are two arbitrary real constants,
$\varphi = \tan^{-1}(\beta/\alpha)$, and the $i$
in  front of $\beta$ in the left side of the last formula
is needed to make that term hermitian.
Redefining the operators $d$ and $d^{\dagger}$ to absorb the
phase $\varphi$, we reduce the hybridizing combination to either
$d - d^{\dagger}$ or $d + d^{\dagger}$.
In other words, we can always choose a proper definition
for $d$ and $d^{\dagger}$  so that only one
of $d \pm d^{\dagger}$ hybridizes
with $\psi_{sf}(0)-\psi_{sf}^{\dagger}(0)$.
But we can only perform  phase absorption once.
A redefinition of  $d$ and $d^{\dagger}$ to
absorb a second phase to eliminate
$ [\psi_{sf}(0)+\psi_{sf}^{\dagger}(0)]
(d+d^{\dagger}) $ from the
stageered field coupling terms in
(\ref{HFP})+(\ref{HSTAG_MORE}) is not possible
without spoiling the simple
hybridization form of the fixed point Hamiltonian~(\ref{HFP}).

 When  the  particle-hole symmetry  is broken,
 another  relevant operator
 becomes allowed,
 as can be seen from Table~\ref{dim12}.
This is the dimension 1/2 operator in (\ref{hphb}).
There are also two dimension 1
operators breaking only
 particle-hole
 symmetry.
{}From Table~\ref{dim1}, they are
\begin{equation}
H'_{phb} = V \left[ \psi_{sf}(0)-\psi_{sf}^{\dagger}(0) \right]
 \left[ \psi_{f}(0) +\psi_{f}^{\dagger}(0) \right]
+ i \alpha_{v} V   \left[ \psi_{f}(0)+\psi_{f}^{\dagger}(0) \right]
 (d+d^{\dagger}) ,
	\label{HPHB_MORE}
\end{equation}
where $\alpha_{v}$ is a dimensionless coefficient
depending on the original coupling constants of (\ref{hamil}).
The first term in (\ref{HPHB_MORE})
is the original particle-hole symmetry breaking term
(\ref{h2_3}), surviving the first order  projection.
The second term is a generated one from high orders
and cannot be eliminated by  a simple phase absorption in
$d$ and $d^{\dagger}$
for the same reason of the last paragraph.

Summarizing this section,
 (\ref{heff})+(\ref{HSTAG_MORE})+(\ref{HPHB_MORE})
constitutes the most general effective Hamiltonian
for the solution region of Figure~\ref{phase_diag}, even
allowing particle-hole symmetry breaking.
What are omitted up to dimension 3/2
only include:
\begin{itemize}
\item
  A dimension 1 operator
$ \left[ \psi_{sf}(0)+\psi_{sf}^{\dagger}(0) \right]
 \left[ \psi_{f}(0) +\psi_{f}^{\dagger}(0) \right] $,
which could couple to the staggered field $h_{s}$
but breaks  the particle-hole symmetry.
Thus, the coefficient of this operator must be proportional to
the particle-hole symmetry breaking potential $V$.
Close to the critical point, we expect this coefficient
to be significantly suppressed.
This term should be even less important than
those in (\ref{HSTAG_MORE}).
\item
 Two dimension 3/2 operators breaking  only particle-hole
symmetry, as can be seen from  Table~\ref{dim32}.
They could appear as additional irrelevant operators
in the effective Hamiltonian.
Again, we expect they are significantly suppressed
close to the critical point.
\item
  Several dimension 3/2 operators which could couple to
the uniform or staggered magnetic fields.
Their contributions to the susceptibilities  vanish
according to high powers of temperature as $T \rightarrow 0$.
\end{itemize}
It is worth
 pointing out that up to dimension 3/2
 the number of allowed operators around the critical point and
 their dimensions are in complete agreement with the conformal
 theory results~\cite{affl94}.

  \section{Low energy thermodynamics }  \label{low-T-prop}

In this section, we shall calculate low energy thermodynamic
properties of the effective Hamiltonian~(\ref{heff})
for the solution region of Figure~\ref{phase_diag}.
The marginal operators (\ref{HSTAG_MORE})
will be considered in  appendix~\ref{more_phb}, where we shall
show that their effect
is to slightly renormalize the Kondo  and crossover temperatures.
The contribution to the staggered susceptibility from
the marginal operators (\ref{HSTAG_MORE})
will be considered in appendix~\ref{more_chis}
and  shown to be negligible.
The way we shall adopt to carry out calculations
is to  represent the partition function
as a path integral in which every fermion operator
becomes a Grassmann variable. Then we perform linear transformations
on the Grassmann variables to bring the action to a
diagonal form.

The partition function in the path integral formalism can be written as
\begin{eqnarray}
Z &=& \int {\cal D}[\psi_{sf}, \bar{\psi}_{sf}, \psi_{f}, \bar{\psi}_{f},
d, \bar{d}]
\, e^{-\int^{\beta}_{0} d\tau ( {\cal L}_{0}
+ H_{eff} ) }   ,
	\label{z1}      \\
{\cal L}_{0} &=& \int^{\infty}_{-\infty} dx \left[ \bar{\psi}_{sf}(x)
\partial_{\tau} \psi_{sf}(x) + \bar{\psi}_{f}(x)
\partial_{\tau} \psi_{f}(x) \right]
+ \bar{d} \partial_{\tau} d ,
\end{eqnarray}
where $ H_{eff}$ is  the effective Hamiltonian given by
(\ref{heff}).
By making  linear transformations to new Grassmann variables,
\begin{eqnarray}
 a_{sf} &=& \frac{1}{\sqrt{2}} ( \psi_{sf} + \bar{\psi}_{sf} ),
\hspace{.3in}
b_{sf} = -\frac{i}{\sqrt{2}} ( \psi_{sf} - \bar{\psi}_{sf} ),   \\
 a_{f} &=& \frac{1}{\sqrt{2}} ( \psi_{f} + \bar{\psi}_{f} ),
\hspace{.3in}
b_{f} = -\frac{i}{\sqrt{2}} ( \psi_{f} - \bar{\psi}_{f} ),   \\
 a &=& \frac{1}{\sqrt{2}} ( d + \bar{d} ),
\hspace{.3in}
b = -\frac{i}{\sqrt{2}} ( d - \bar{d} ),
\end{eqnarray}
we write
the total Lagrangian  in (\ref{z1}) as
\begin{eqnarray}
 {\cal L} &=& {\cal L}_{0} + H_{eff}
	= {\cal L}_{1}(a_{sf})
+  {\cal L}_{2}(b_{sf})
+ {\cal L}_{3}(a_{f})
+ {\cal L}_{4}(b_{f})
+{\cal L}_{loc}(a,b)    ,               \\
	{\cal L}_{1}(a_{sf})
&=&  \frac{1}{2}\int^{\infty}_{-\infty} dx \;
 a_{sf}(\tau,x) (\partial_{\tau} -i v_{F} \partial_{x} ) a_{sf}(\tau,x)
+  2 i \alpha_{s} h_{s}    a_{sf}(\tau,0) b(\tau) , \label{l1}  \\
{\cal L}_{2}(b_{sf})
&=&  \frac{1}{2} \int^{\infty}_{-\infty} dx \;
 b_{sf}(\tau,x) (\partial_{\tau} -i v_{F} \partial_{x} ) b_{sf}(\tau,x)
	\nonumber       \\
  & & +  2 i  v_{F}  \left[ g_{0} \,  b_{sf}(\tau,0)   a(\tau)
+   g_{1} b(\tau) \partial_{x}  b_{sf}(\tau,0)  \right]  ,      \\
	{\cal L}_{3}(a_{f})
&=&  \frac{1}{2} \int^{\infty}_{-\infty} dx \;
 a_{f}(\tau,x) (\partial_{\tau} -i v_{F} \partial_{x} ) a_{f}(\tau,x)
+  2 i  \widetilde{V} a_{f}(\tau,0) b(\tau)  ,  \\
	{\cal L}_{4}(b_{f})
&=&  \frac{1}{2} \int^{\infty}_{-\infty} dx \;
 b_{f}(\tau,x) (\partial_{\tau} -i v_{F} \partial_{x} ) b_{f}(\tau,x)
+  2 i  \alpha_{u} h_{u} a_{sf}(\tau,0) b_{f}(\tau,0)  ,        \\
	{\cal L}_{loc}(a,b)
&=&   \frac{1}{2} \left[ a(\tau) \partial_{\tau} a(\tau)
+ b(\tau) \partial_{\tau} b(\tau) \right]
+ i  \, \delta K \; a(\tau)  b(\tau) ,          \label{llocal}
\end{eqnarray}
where
\begin{equation}
 \delta K =  K_{c} - \frac{1}{2} (K_{z}+K_{\perp}) .    \label{def_dK}
\end{equation}
Introducing the Fourier transformation,
\begin{equation}
	a_{sf}(\tau,x) = \frac{1}{\beta} \sum_{n}
\int^{\infty}_{-\infty} \frac{d k}{2\pi}
a_{sf}(\omega_{n} , k) \, e^{-i\omega_{n}\tau + i kx}  ,
\end{equation}
and similar transformations for the other Grassmann variables,
we can write the actions corresponding to the Lagrangians~(\ref{l1})
to (\ref{llocal}) as
\begin{eqnarray}
 & & {\cal S}_{1}(a_{sf}) =
  \sum_{n} \int^{\infty}_{-\infty} \frac{d k}{2\pi}
\left[ -\frac{1}{2}
(i\omega_{n} - v_{F} k)  a_{sf}(-\omega_{n} , -k) a_{sf}(\omega_{n} , k)
	\right.  \nonumber      \\
& & \hspace{.5in}       \left.
+ 2 i  \alpha_{s} h_{s} \,
 a_{sf}(\omega_{n} , k)  b(-\omega_{n}) \right]  , \label{S1}  \\
 & & {\cal S}_{2}(b_{sf}) =
  \sum_{n} \int^{\infty}_{-\infty} \frac{d k}{2\pi}
\left\{ -\frac{1}{2}
(i\omega_{n} - v_{F} k)  b_{sf}(-\omega_{n} , -k) b_{sf}(\omega_{n} , k)
	\right.         \nonumber               \\
 & & \hspace{.5in}      \left.
- 2 i v_{F} \left[ g_{0} a(-\omega_{n})
 - i g_{1} \, k \,  b(-\omega_{n})
  \right]  b_{sf}(\omega_{n} , k)   \right\}  ,         \label{S2}  \\
 & &  {\cal S}_{3}(a_{f}) =
  \sum_{n} \int^{\infty}_{-\infty} \frac{d k}{2\pi}
\left[ -\frac{1}{2}
(i\omega_{n} - v_{F} k)  a_{f}(-\omega_{n} , -k) a_{f}(\omega_{n} , k)
+  2 i  \widetilde{V}  a_{f}(-\omega_{n} , -k)  b(\omega_{n})
\right] ,               \label{S3}      \\
 & &  {\cal S}_{4}(b_{f}) =
  \sum_{n} \int^{\infty}_{-\infty} \frac{d k}{2\pi}
\left[ -\frac{1}{2}
(i\omega_{n} - v_{F} k)  b_{f}(-\omega_{n} , -k) b_{f}(\omega_{n} , k)
	\right.  \nonumber      \\
& & \hspace{.5in}  \left.
 + 2i  \alpha_{u} h_{u} \,
a_{sf}(-\omega_{n} , -k) b_{f}(\omega_{n} , k)
\right] ,               \label{S4}              \\
 & & {\cal S}_{loc}(a,b) =
  \sum_{n} \left\{ - \frac{i\omega_{n} }{2} \left[
a(-\omega_{n}) a(\omega_{n}) +  b(-\omega_{n}) b(\omega_{n}) \right]
+ i  \, \delta K \;  a(-\omega_{n})  b(\omega_{n})
\right\}  .             \label{slocal}
\end{eqnarray}
The uniform magnetic field term in (\ref{S4}) is an exactly marginal
operator.
This is most easily seen by setting $h_{s} = 0$ and combining
(\ref{S1}) with (\ref{S4}). From the real Grassmann
variables $a_{sf}$ and $b_{f}$, we can make a linear
transformation to
\[ \psi=\frac{1}{\sqrt{2}} ( a_{sf} + i b_{f} ),
\hspace{.3in}
\bar{\psi}=\frac{1}{\sqrt{2}} ( a_{sf} - i b_{f} )  . \]
The Grassmann variables $ \psi $ and $\bar{\psi}$
correspond to the usual fermion annihilation and creation
operators.
 When $h_{s}=0$, (\ref{S1})+(\ref{S4}) is completely
decoupled from the rest of the Hamiltonian
responsible for the critical behavior.
Moreover,
the $h_{u}$ term of (\ref{S4})
is simply a potential scattering term in terms of the
fermions corresponding to $ \psi $ and $\bar{\psi}$,
\begin{equation}
{\cal S}(a_{sf}) + {\cal S}(b_{f})
=\int^{\beta}_{0} d\tau \left[
\int^{\infty}_{-\infty} dx \bar{\psi}(x)
(\partial_{\tau} - i v_{F} \partial_{x} ) \psi(x)
+ 2 \alpha_{u} h_{u} \bar{\psi}(0) \psi(0) \right] .
\end{equation}
Thus, not only
the uniform susceptibility is well behaved
but also applying a uniform external magnetic field has
negligible effect on the physical behavior
of the system inside the solution region of Figure~\ref{phase_diag}!
{}From now on, we shall set $h_{u}=0$ and
drop (\ref{S4}) from further discussion.

We can diagonalize (\ref{S1}) to (\ref{S3}) simply by shifting
the Grassmann variables corresponding to the extended
degrees of freedom,
\begin{eqnarray}
\widetilde{a}_{sf}(\omega_{n} , k)      &=&
a_{sf}(\omega_{n} , k)
- \frac{ 2 i  \alpha_{s} h_{s}  }{i\omega_{n} - v_{F} k }
b( \omega_{n}) ,
		\\
\widetilde{b}_{sf}(\omega_{n} , k)      &=&
b_{sf}(\omega_{n} , k)
- \frac{ 2 i v_{F}  }{i\omega_{n} - v_{F} k }
\left[ g_{0} a( \omega_{n}) + i g_{1} k  b( \omega_{n}) \right] ,
		\label{shift_bsf}               \\
\widetilde{a}_{f}(\omega_{n} , k)      &=&
a_{f}(\omega_{n} , k)
- \frac{ 2 i  \widetilde{V}   }{i\omega_{n} - v_{F} k }
b( \omega_{n}) .
		\label{shift_af}
\end{eqnarray}
Upon inserting the results for the following   integrals,
\begin{eqnarray}
\int^{\infty}_{-\infty} \frac{d k}{2\pi}
 \frac{1}{i\omega_{n} - v_{F} k }
 &=&  - \frac{ i \, {\rm sgn}\omega_{n} }{ 2 v_{F} } ,
	\label{integ1}  \\
\int^{\Lambda}_{-\Lambda} \frac{d k}{2\pi}
 \frac{k}{i\omega_{n} - v_{F} k }
 &=&    - \frac{1}{v_{F}^{2}} \left(
 \frac{ v_{F} \Lambda}{\pi } - \frac{ |\omega_{n}| }{2} \right) ,
	\label{integ2}  \\
\int^{\Lambda}_{-\Lambda} \frac{d k}{2\pi}
 \frac{k^{2}}{i\omega_{n} - v_{F} k }
 &=&  - \frac{i \omega_{n} }{ v_{F}^{3} } \left(
\frac{ v_{F} \Lambda }{\pi} - \frac{ |\omega_{n} | }{2} \right) ,
	\label{integ3}
\end{eqnarray}
where $\Lambda$ is the ultraviolet cutoff,
the actions~(\ref{S1}) to (\ref{S3}) become
\begin{eqnarray}
 {\cal S}_{1}(a_{sf}) &=&  -
\sum_{n} \int^{\infty}_{-\infty} \frac{d k}{4\pi}
 (i\omega_{n} - v_{F} k)
\widetilde{a}_{sf}(-\omega_{n} , -k) \widetilde{a}_{sf}(\omega_{n} , k)
	\nonumber               \\
 & & \hspace{.5in}
- \frac{ i ( \alpha_{s} h_{s} )^{2} }{v_{F}} \sum_{n} {\rm sgn}\omega_{n} \,
 b( -\omega_{n})  b( \omega_{n}) ,       \label{S1_1}       \\
 {\cal S}_{2}(b_{sf}) &=&  -
\sum_{n} \int^{\infty}_{-\infty} \frac{d k}{4\pi}
 (i\omega_{n} - v_{F} k)
\widetilde{b}_{sf}(-\omega_{n} , -k) \widetilde{b}_{sf}(\omega_{n} , k)
 -i v_{F} g_{0}^{2} \sum_{n} {\rm sgn}\omega_{n}  \,
 a( -\omega_{n})  a( \omega_{n})
		\nonumber               \\
 &  & + 2ig_{0}g_{1}  \sum_{n} |\omega_{n}| a( -\omega_{n}) b( \omega_{n})
+  \frac{ i g_{1}^{2}}{v_{F}}  \sum_{n}  \omega_{n}
\left( |\omega_{n}| - \frac{ 2 v_{F} \Lambda }{\pi} \right)
 b( -\omega_{n}) b( \omega_{n})
		\nonumber               \\
 & & -  \frac{4 i v_{F} g_{0} g_{1}  \Lambda }{\pi}  \sum_{n}
 a( -\omega_{n}) b( \omega_{n}) ,               \label{S2_1}  \\
 {\cal S}_{3}(a_{f}) &=&  -
\sum_{n} \int^{\infty}_{-\infty} \frac{d k}{4\pi}
 (i\omega_{n} - v_{F} k)
\widetilde{a}_{f}(-\omega_{n} , -k) \widetilde{a}_{f}(\omega_{n} , k)
- \frac{ i \widetilde{V}^{2} }{v_{F}} \sum_{n} {\rm sgn}\omega_{n}  \,
 b( -\omega_{n})  b( \omega_{n}) .              \label{S3_1}
\end{eqnarray}
The last term in (\ref{S2_1}) is a correction to the
relevant operator $\delta K$ term of (\ref{slocal})
and can be absorbed into the critical value of
the RKKY interaction $K_{c}$.
Collecting the local terms containing Grassmann variables $a$, $b$
from (\ref{S1_1}) to (\ref{S3_1})
and combining  them with (\ref{slocal}), we obtain
the effective local action,
\begin{eqnarray}
{\cal S}_{loc}^{eff} &=&
 - i \, \sum_{n>0} \left(  a( -\omega_{n}),  b( -\omega_{n}) \right)
		\nonumber       \\
	& \times &
  \left( \begin{array}{cc}
 |\omega_{n}| + 2v_{F}g_{0}^{2}   &
-( \delta K + 2 g_{0} g_{1} |\omega_{n}|  )                 \\
\delta K + 2 g_{0} g_{1} |\omega_{n}|     \; \;         &
 \; |\omega_{n}| / Z_{b}
+ 2  \left( \alpha_{s}^{2} h_{s}^{2}
+  \widetilde{V}^{2}
- g_{1}^{2} \omega_{n}^{2} \right)/ v_{F}      \\
\end{array}     \right)
\left(  \begin{array}{c}
	 a( \omega_{n}) \\   b( \omega_{n})
	\end{array}     \right)  .      \label{seff_loc}
\end{eqnarray}
The factor $Z_{b}$ is defined by
\begin{equation}
  Z_{b} = \frac{1}{ 1+ 4 g_{1}^{2}\Lambda/\pi }  , \label{def_zb}
\end{equation}
and  can be interpreted as
the wave function renormalization
factor for  the Grassmann variable $b$(or Majorana fermion).
The $\omega_{n}^{2}$  term in the matrix element of (\ref{seff_loc})
can be safely neglected since it is highly irrelevant
and satisfies $ g_{1} \omega_{n}^{2} / v_{F} \ll | \omega_{n} | $
for the whole energy range of practical interest.
All interesting thermodynamics is contained in (\ref{seff_loc}).
{}From (\ref{seff_loc})  we obtain the free energy shift
due to the local interactions,
\begin{eqnarray}
F(T,h_{s}) &=& -T \ln 2 - T \sum_{n > 0} \ln
\left\{
  (\delta K + 2 g_{0} g_{1} |\omega_{n}| )^{2}
		\right.         \nonumber       \\
 & & \left.
 +  \left(|\omega_{n}|+2v_{F}g_{0}^{2} \right)
\left[ \frac{ |\omega_{n}| }{ Z_{b} }
+ \frac{2 }{v_{F}} \left( \alpha_{s}^{2} h_{s}^{2}
+  \widetilde{V}^{2} \right)    \right]
  \right\} ,
\end{eqnarray}
where  $-T \ln 2 $ is the entropy of two degenerate
impurity spin states.
Defining several convenience notations,
\begin{eqnarray}
T_{K} & =&  2 v_{F} g_{0}^{2} + 2 Z_{b}  \left(
\frac{ \widetilde{V}^{2} }{v_{F} }
+ 2 g_{0} g_{1} \, \delta K \right) ,        \label{def_tk}      \\
T_{c} &=& \frac{ Z_{b} \left[
4g_{0}^{2} \widetilde{V}^{2} + (\delta K)^{2} \right] }{T_{K}}  ,
			\label{def_tc}   \\
\widetilde{\alpha}_{s} &=& 2 g_{0} \alpha_{s} \sqrt{ Z_{b} },
		\label{def_tilde_as}
\end{eqnarray}
we can recast the free energy in a very simple form,
\begin{equation}
F(T, h_{s} ) = - T \ln 2
+ \int^{\infty}_{-\infty} \frac{ d \omega}{2\pi}
\frac{1}{e^{\beta \omega} + 1}
\tan^{-1} \left\{ \frac{ \omega \left[ T_{K}
+  \widetilde{\alpha}_{s}^{2} h_{s}^{2}
	/ (2v_{F} g_{0}^{2})  \right] }
{\omega^{2}- T_{c}T_{K} - \widetilde{\alpha}_{s}^{2} h_{s}^{2}  }
\right\} .              \label{fe1}
\end{equation}
The roles of the parameters in (\ref{fe1}) can be read off.
 $T_{K}$ is the fundamental energy scale of
the problem and should be identified as the Kondo temperature.
We note $T_{K} \simeq 2 v_{F} g_{0}^{2}$,
as can be seen from (\ref{def_tk}).
$T_{c}$   vanishes approaching
the critical point,
and satisfies $T_{c} \ll T_{K}$ inside the solution region
of Figure~\ref{phase_diag}.  The same $T_{c}$
defines the crossover
energy scale above which the behavior of the system is
controlled by the critical point. Below $T_{c}$,
it is controlled by the Fermi-liquid fixed point.
Accompanying the staggered magnetic field is an involved coefficient
$\widetilde{\alpha}_{s}$  because $h_{s}$
couples to an unconserved operator. Because of this factor,
it is not possible
to define a universal Wilson ratio from the staggered susceptibility.

For all practical purposes,  the $h_{s}^{2}$ term in the numerator
inside  $\tan^{-1}$
in (\ref{fe1}) can be dropped since it only shifts
$T_{K}^{2}$ to $T_{K}^{2}+\widetilde{\alpha}_{s}^{2} h_{s}^{2}$.
After some rearrangement, we finally obtain
\begin{equation}
F(T, h_{s})  =
-T \ln 2
 - \int^{\infty}_{0} \frac{ d \omega}{ 2 \pi }
 \tanh \left(\frac{\beta\omega}{2}\right)
\tan^{-1} \left( \frac{ \omega  T_{K}   }
{ \omega^{2}  - T_{c}T_{K} - \widetilde{\alpha}_{s}^{2} h_{s}^{2}  }
\right)   .            \label{fe2}
\end{equation}
This nice looking
expression gives us the complete crossover functions for the
specific heat and staggered susceptibility.

\subsection{Specific Heat}

Setting $h_{s}=0$ in (\ref{fe2})
and performing some minor manipulation, we obtain
\begin{equation}
F(T) - F(0) =
-T \ln 2 + T \int^{\infty}_{0} \frac{dx}{\pi}
\frac{1}{e^{x} + 1}
\tan^{-1} \left( \frac{ x \beta T_{K} }
{x^{2} - \beta^{2} T_{c}T_{K}    } \right) .    \label{fe3}
\end{equation}
Two limiting behaviors follow immediately.
At $T_{c} \ll T \ll T_{K} $,
\begin{equation}
F(T) - F(0) =
-T \ln 2 + T \int^{\infty}_{0} \frac{dx}{\pi}
\frac{1}{e^{x} + 1}
\left( \frac{\pi}{2} - \frac{x}{\beta T_{K}} \right)
= - \frac{T}{2} \ln 2 - \frac{\pi}{12}
\frac{T^{2}}{ T_{K} } .             \label{c_lim1}
\end{equation}
We see a residual entropy $( \ln 2)/2$, reduced from
the original $\ln 2$.
At  $T \ll T_{c} \ll T_{K}$,
\begin{equation}
F(T) - F(0) =
-T \ln 2 + T \int^{\infty}_{0} \frac{dx}{\pi}
\frac{1}{e^{x} + 1}
\left( \pi -
\frac{ x  }{\beta T_{c} }   \right)
= - \frac{\pi}{12} \frac{ 1 }{ T_{c} } T^{2} .
	\label{c_lim2}
\end{equation}
The limiting behaviors of the specific heat are obvious from
(\ref{c_lim1}) and (\ref{c_lim2}),
\begin{equation}
\frac{C(T, T_{c})}{T} =
\left\{  \begin{array}{ll}
	\frac{ \pi }{ 6 T_{K} } , &  T \gg T_{c} ,             \\
	 \frac{ \pi }{  6 T_{c} }  ,
	&   T_{c} \gg T  .          \\
	 \end{array}         \right.            \label{ct_lim}
\end{equation}
The general crossover function for the specific heat
is obtained from (\ref{fe3}),
\begin{equation}
\frac{C(T)}{T} =     2 T_{K}^{2}  \beta^{4}
\int^{\infty}_{0} \frac{dx}{\pi}   \frac{x}{e^{x}+1}
\frac{ x^{4} T_{K}
- T_{c} ( x^{2} - \beta^{2} T_{c}T_{K} )
( 3 x^{2} + \beta^{2} T_{c}T_{K} )  }{ \left[
(x^{2}-\beta^{2}T_{c}T_{K} )^{2} + (x \beta T_{K} )^{2}
\right]^{2} }.                \label{exp_ccross}
\end{equation}
This crossover function is plotted in Figure~\ref{c_cross}.
We also plot
the crossover function for the entropy~(\ref{fe3})
in Figure~\ref{entrop_cross}, which may be more instructive.

\subsection{Staggered Susceptibility}

{}From (\ref{fe2}), the staggered susceptibility is given by
\begin{equation}
\chi_{s}(T, T_{c}) =
- \left[
\frac{\partial^{2}}{\partial h_{s}^{2}} F(T,h_{s}) \right]_{h_{s}=0} =
\widetilde{\alpha}_{s}^{2} \int^{\infty}_{0}
\frac{d\omega}{\pi} \tanh\left(\frac{\beta\omega}{2}\right)
\frac{\omega T_{K} }{ (T_{c} T_{K} -\omega^{2})^{2}
+(\omega T_{K})^{2} } .            \label{chis_cross}
\end{equation}
The limiting behaviors are found to be
\begin{equation}
\chi_{s}(T, T_{c}) =
\left\{  \begin{array}{ll}
	- \frac{ \widetilde{\alpha}_{s}^{2} }{ \pi T_{K} }
		\ln T , &  T \gg T_{c} ,             \\
	 - \frac{ \widetilde{\alpha}_{s}^{2} }{ \pi T_{K} }
		\ln T_{c} , &   T  \ll T_{c}  .          \\
	 \end{array}         \right.            \label{chis_lim}
\end{equation}
The crossover behavior for
 the function (\ref{chis_cross}) is plotted in
Figure~\ref{fig_chis}.

\subsection{Impurity Spin Correlation
$\langle \vec{S}_{1} \cdot \vec{S}_{2} \rangle $  }

Projecting the operator $(\vec{S}_{+})^{2}$
 onto the doublet, we find
 \begin{equation}
\hat{Q}  \vec{S}_{+}^{2} \hat{Q}
 =  \hat{Q} (S_{+}^{x})^{2} \hat{Q} +
  \hat{Q} (S_{+}^{y})^{2} \hat{Q} +
  \hat{Q} (S_{+}^{z})^{2} \hat{Q}
  = 2 d d^{\dagger}  .
\end{equation}
 Thus, the calculation of the impurity spin correlation is
reduced to evaluating $\langle d d^{\dagger} \rangle$.
{}From (\ref{seff_loc}), we have
\begin{equation}
 \langle d d^{\dagger} \rangle - \frac{1}{2}
=   i \langle b a \rangle
=  T \sum_{n}
\frac{ \delta K + 2 g_{0} g_{1} | \omega_{n} | }{
 (\delta K + 2 g_{0} g_{1} |\omega_{n}| )^{2}
 +  \left(|\omega_{n}|+ T_{K} \right)
\left( |\omega_{n}| /Z_{b}
+  2 \widetilde{V}^{2} / v_{F} \right)   }  .
\end{equation}
The impurity spin correlation is
\begin{equation}
\langle \vec{S}_{1} \cdot \vec{S}_{2} \rangle
= -\frac{1}{4} + \frac{Z_{b} }{\beta}
\sum_{n} \frac{ \delta K + 2 g_{0} g_{1} | \omega_{n} | }{
\omega_{n}^{2} + T_{c}T_{K} + |\omega_{n}|  T_{K}  } .
	\label{cor_s1s2}
\end{equation}

At the critical point, $\delta K = T_{c} =0$, and as $T \rightarrow 0$,
\begin{equation}
\langle \vec{S}_{1} \cdot \vec{S}_{2} \rangle
\simeq  - \frac{1}{4} + Z_{b}
\int^{v_{F} \Lambda }_{0} \frac{d\omega}{\pi}
\frac{ 2g_{0}g_{1} }{\omega + T_{K} }
\simeq -\frac{1}{4} +
\frac{2  g_{0} g_{1} Z_{b} }{ \pi }
\ln \frac{ v_{F} \Lambda }{  T_{K} } .
\end{equation}
The leading irrelevant operator induces a small
non-universal correction
to the impurity spin correlation.
Although the fixed point itself(the ground state)
has an extra
symmetry between the two states of the doublet,
$d \leftrightarrow d^{\dagger}$, which
implies $ \langle \vec{S}_{1} \cdot \vec{S}_{2} \rangle
= - 1/4$, it is broken by
the leading irrelevant operator, as can be seen from (\ref{HPERT}).

We can also calculate the  slope of  the impurity spin correlation
with respect to the RKKY interaction.
{}From (\ref{cor_s1s2}), we find
\begin{eqnarray}
\frac{\partial}{\partial \, (\delta K)}
\langle \vec{S}_{1} \cdot \vec{S}_{2} \rangle
&=&  \frac{ Z_{b} }{\beta}
\sum_{n} \frac{1}{ \omega_{n}^{2} +  |\omega_{n}|  T_{K}
+ T_{c}T_{K} }
		\nonumber               \\
&=&  Z_{b}
\int^{\infty}_{0} \frac{d\omega}{\pi}
 \tanh\left(\frac{\beta\omega}{2}\right)
\frac{ \omega  T_{K} }{  (T_{c}T_{K} -\omega^{2})^{2} +
( \omega  T_{K} )^{2} } .               \label{slope_s1s2}
\end{eqnarray}
Comparing the last expression with (\ref{chis_cross}), we find
\begin{equation}
\frac{\partial}{\partial \, (\delta K)}
\langle \vec{S}_{1} \cdot \vec{S}_{2} \rangle
= const. \times  \chi_{s}(T, T_{c}) .           \label{chis_s1s2}
\end{equation}
In particular, they should have the same limiting behaviors
as given by (\ref{chis_lim}).
Since
\begin{equation}
\frac{\partial}{\partial \, (\delta K)}
\langle \vec{S}_{1} \cdot \vec{S}_{2} \rangle
\sim \int^{\beta}_{0} d\tau
\langle \hat{T} (\vec{S}_{1} \cdot \vec{S}_{2})(\tau) \;
(\vec{S}_{1} \cdot \vec{S}_{2})(0) \rangle  ,  \label{s1s2_2}
\end{equation}
the result~(\ref{chis_s1s2}) should not be too surprising.

\section{Comparison with other results and universality} \label{compare}

 There are two kinds of asymptotically exact limiting results
 with which we can compare our solution.
 These are the conformal field theory
 results at $T \gg T_{c}$
and the numerical renormalization group
 results at $T=0$. First of all, we would like to emphasize
 that all our results are also asymptotically exact
 up to some numerical
coefficients $g_{0}$, $g_{1}$ and $\widetilde{V}$
 in (\ref{heff}).
 Or eventually, the possible uncertainty
 boils down to the two basic energy scales
 $T_{K}$ and $T_{c}$, which we cannot
 determine  exactly in terms of the initial parameters of
 the original Hamiltonian~(\ref{h1}).

  In order to make a comparison, we first need to answer
  the questions that whether or not
  the critical point we  have studied is the same one,
  and whether or not the spin anisotropy we have introduced
  in (\ref{h1}) is irrelevant.
  The answer to both questions is a convincing yes,
  if not rigorous.
   A detailed comparison of the finite size spectrum of the
   critical point between
   the conformal field theory
 and numerical renormalization group approaches has
 been made~\cite{affl94}.
   Excellent agreement
   has been found which
   implies the same critical point in those two approaches.
   Thus, we shall take the agreement between our results
   and that obtained from either one of those two approaches
   as a positive evidence.
   The conformal field
   theory  tells us that there is only one non-Fermi-liquid
   fixed point, {\it i.e.} conformally
   invariant boundary condition~\cite{affl94}.
   This is supported by the failure of finding other critical
   points in our approach by considering
   other impurity spin states
   as the lowest degenerate levels than the
   doublet of Figure~\ref{loc_lev}.
 The strongest evidence for the universality
 of the critical point
 is the exactly same operator content
 around the critical point we find in our approach and
 in the conformal field theory approach. That is we have the
 same number of operators with  the same symmetry and same dimension.
 Specifically, there is one dimension 1/2 relevant operator
 and one dimension 3/2 leading irrelevant operator in the presence
 of the particle-hole symmetry, as can be seen from (\ref{HFP}) and
(\ref{HPERT}).
 Breaking the particle-hole symmetry introduces another
 dimension 1/2 relevant operator, as can be seen from (\ref{hphb}).
 Furthermore, the  dimension 3/2
 leading irrelevant operator in the conformal
 field theory is a descendent of the relevant operator.
 In our approach, we  consistently find that the leading irrelevant
 operator contains the spatial derivative $\partial_{x}$.
This is the crucial difference from the two-channel Kondo problem,
resulting in different low temperature behavior for the
specific heat(see (\ref{ct_lim}) and reference~\cite{emery92}).
While in both cases there is
 a dimension 3/2 leading
irrelevant operator, only the one
at the critical point of the two-impurity Kondo problem
contains $\partial_{x}$.
 As to the spin anisotropy, it is found
  in the conformal field theory approach  that
 a small spin anisotropic
 perturbation around the critical point is  irrelevant~\cite{affl94}.
 Although this does not prove the irrelevance of the
 spin anisotropy introduced in our approach
 because the introduced anisotropy is not small,
 it does point to the right
 direction.
It is worthwhile recalling
that the spin anisotropy is irrelevant for all kinds
of one-impurity Kondo problem,
 including the exactly screened~\cite{toul70}
and overscreened  cases~\cite{affl92,georges94}.
Leaving aside specifics of the
 two-impurity Kondo model,
 these early experiences in related Kondo problems
 give us considerable confidence that the critical point
 of the two-impurity model
 that has  been studied from different approaches  belongs to
 the same universality class.

 Since the behavior of the system above
 $T_{c}$ and the way the system flows
to the stable Fermi-liquid fixed point below $T_{c}$ are all
 governed by the critical point,
the universality of the critical point
 also implies the universal behavior
inside the solution region of Figure~\ref{phase_diag},
 as well as all crossover functions.  In particular,
 the crossover functions we have derived for the specific heat
 and staggered susceptibility are expected to be universal.
 For a comparison of the results from different approaches,
 the only freedom left is to match the two basic energy scales
 $T_{K}$ and $T_{c}$. For the staggered susceptibility,
 or any other response function of a non-conserved operator,
 there may also be an undetermined
overall constant prefactor.

At the critical point $T_{c}=0$, or more generally in the limit
 $T \gg T_{c}$,
the critical properties of all thermodynamic quantities
as a function of the temperature that
we have calculated in the last section
completely agree with the conformal field theory results,
as expected on the grounds of the same operator content.
These include: the residual entropy $(\ln 2)/2$, linear
specific heat, $\ln T$ singularity in the staggered susceptibility,
constant uniform susceptibility, and $\ln T$ singularity
in the correlation function of the composite
operator $\vec{S}_{1} \cdot \vec{S}_{2}$, as can be seen
from (\ref{slope_s1s2}) and (\ref{s1s2_2}).
The complete agreement of the critical behavior further ensures us
the  universality  of the critical point.

As to compare with the numerical
renormalization group results at $T=0$, we first note that
the empirical observation~\cite{jones}
 of an additional hidden symmetry between the
singlet and triplet impurity
spin states at the critical point becomes
crystal clear after our identification of the critical point,
as can be seen from Figure~\ref{loc_lev}. So is its consequence
about the value of the impurity spin correlation at the
critical point, $\langle \vec{S}_{1} \cdot \vec{S}_{2} \rangle =-1/4$.
However,
this hidden symmetry is broken by the  leading irrelevant
operator,  as we have noted before. This is similar to the
well known dynamical symmetry breaking effect.
In consistency with the numerical renormalization group result,
we also find that the linear coefficient of the
specific heat diverges quadratically in $\delta K$ on the
particle-hole symmetric axis, as can be seen from (\ref{ct_lim}).
Our result~(\ref{chis_s1s2}), that  the slope
of the impurity spin correlation with respect to
the variation of RKKY interaction is logarithmically divergent,
is broadly consistent with the numerical renormalization
group result which
also found it singular.

The only disagreement with
the reported numerical renormalization group results is the behavior of
the staggered susceptibility at $T=0$. While
$\chi_{s}(T=0,T_{c}) \sim 1/T_{c}$ has be claimed, we only find
$\chi_{s}(T=0,T_{c}) \sim \ln T_{c}$, as can be seen from (\ref{chis_lim}).
Noting that the other limiting behavior of (\ref{chis_lim}),
$\chi_{s}(T,T_{c}) \sim \ln T$ at $T \gg T_{c}$,
is not disputed.
Even if one takes a cautious view about the numerically fitted
critical exponent 2 for the staggered susceptibility,
{\it i.e.} $\chi_{s} \sim (\delta K)^{-2}$,
the original numerical divergence seems to us much
stronger than a logarithmic singularity.
The reason for this discrepancy is unknown at this moment.
But at least the easy explanation of differently adopted
definitions for the staggered susceptibility is unlikely.
In this paper, we only couple the staggered field
$h_{s}$ to $S_{-}^{z}$ in (\ref{h01}). One could also couple
$h_{s}$ to
$[\psi_{1}^{\dagger}(0)\sigma^{z}\psi_{1}(0)
  -\psi_{2}^{\dagger}(0)\sigma^{z}\psi_{2}(0)]/2$ in (\ref{h01}),
or even to $\int dx [\psi_{1}^{\dagger}(x)\sigma^{z}\psi_{1}(x)
  -\psi_{2}^{\dagger}(x)\sigma^{z}\psi_{2}(x)]/2$.
In any case, the contributions to the staggered susceptibility
after subtracting out  the free Fermi sea contribution
should only come from the local operators which are odd
under parity and $\pi$ rotation around $x$-axis.
Since we only have one such relevant operator,
as can be seen from Table~\ref{dim12}, we do not expect qualitative change
of the behavior of the staggered susceptibility
as a result of different definitions.
A careful  reexamination in the numerical renormalization
group approach  should be very helpful
to clarify this point.

\section{Conclusion}    \label{conclusion}

	We have presented an asymptotically exact solution
for the two-impurity Kondo model for a finite region of
the parameter space
surrounding the critical point, as shown in Figure~\ref{phase_diag}.
We have also derived the
 analytic crossover functions for the specific heat and staggered
 susceptibility.
This solution is made possible by an explicit identification
of the critical point and its underlying physics.
As we have explained in section~\ref{mapping},
the condition for the criticality is the degeneracy
between the two lowest impurity spin states,
 $ ( |\uparrow\downarrow> - \, |\downarrow\uparrow> )/\sqrt{2} $
 and
 $ ( |\uparrow\uparrow> + \, |\downarrow\downarrow> )/\sqrt{2} $.
By varying RKKY interaction across the critical point,
these two levels cross each other.
The  non-Fermi-liquid behavior at the critical  point
is  a consequence
of the fact that in the presence of the
particle-hole symmetry only half  of the degrees of freedom of the doublet
can be compensated by the extended degrees of freedom associated
with the conduction electrons.
Because of the special symmetry of the doublet,
{\it i.e.} one level is an even
triplet and the other is an odd singlet,
the local  degrees of
freedom of the doublet can only
be compensated by the extended spin-flavor degrees of freedom
of the conduction electrons. Out of four species of spinless fermions or
eight species of Majorana fermions associated with all conduction
electron degrees of freedom, only one species of Majorana fermions,
$\psi_{sf}-\psi_{sf}^{\dagger}$, is allowed by the symmetry to
compensate the local degrees of the freedom of
the doublet.
This is the same physics responsible for the non-Fermi-liquid
behavior of the two-channel one-impurity Kondo model.
However, the doublet at the critical point of
 the two-impurity problem has different symmetry
from the simple impurity spin up and down states of the two-channel problem.
Therefore, the operator contents around
the fixed points(one unstable, the
other stable) are different.
 The  nearly complete  agreement of our results
  with those derived from the numerical renormalization
group or conformal field theory approaches,
except one limiting behavior of the staggered susceptibility,
convincingly establishes the universality of the critical point.
Thus,  the crossover functions we have derived in
section~\ref{low-T-prop} are also expected to be universal.
The calculation of dynamical correlation functions such as
the conduction electron Green's function is currently
under way.

What have  we learned about the
lattice problem from the study of the two-impurity Kondo model?
An obvious lesson is learned from the striking difference
between the uniform and staggered susceptibilities.
This difference is solely
due to the competition between RKKY interaction and the Kondo
effect. A direct and primitive
translation to the Kondo lattice problem
would be the strong momentum $\vec{q}$ dependence
of the spin susceptibility $\chi''(\omega,\vec{q})$.
As  a result of the competition, we should expect drastically
different enhancement at different momentum
transfer $\vec{q}$. From this perspective,
 the picture of
a periodic array of coherent Kondo scattering centers
for the heavy fermion compounds
is surely oversimplified.
Non-perturbatively incorporating
RKKY interaction into the Kondo effect in the lattice
is an outstanding problem, on which the impact of
the insight from the two-impurity Kondo model
has to be fully realized.

\acknowledgements

The author acknowledges very fruitful discussion with Ian Affleck and
Eugene Wong. This work is supported  by NSERC of Canada.

\appendix
\section{Derivation of (\ref{UH0U}) and (\ref{H1_2}) }
	 Under the transformation  $\hat{U} H \hat{U}^{-1}$,
only two terms in $H_{0}$  are affected. They are
$ h_{u}/(2\pi)  \int^{\infty}_{-\infty}
dx \,   \partial_{x} \Phi_{s}(x) $ and
\begin{equation}
H_{0}^{(s)} = \frac{v_{F}}{2}  \int^{\infty}_{-\infty}
dx \left\{ \Pi_{s}^{2}(x)
+ [ \partial_{x} \phi_{s}(x) ]^{2} \right\} .
\end{equation}
As for $H_{1}$, the transformation  affects the term containing
$ \partial_{x} \Phi_{s}(0) S^{z}_{+} $ apart from eliminating
$\cos\Phi_{s}(0)$ and $\sin\Phi_{s}(0)$.
Using  the mode expansions,
\begin{eqnarray}
\phi_{s}(x) &=& \int^{\infty}_{-\infty} \frac{dp}{2\pi\sqrt{2|p|} }
\left[ \phi_{s}(p) e^{ipx} + \phi_{s}^{\dagger}(p) e^{-ipx} \right]
e^{-\alpha|p|/2} ,         \label{fouri_phi}     \\
\Pi_{s}(x) &=& \int^{\infty}_{-\infty}
\frac{dp \, |p| }{2\pi\sqrt{2|p|} }
\left[-i \phi_{s}(p) e^{ipx}
+ i \phi_{s}^{\dagger}(p) e^{-ipx} \right]
e^{-\alpha|p|/2} ,         \label{fouri_pi}
\end{eqnarray}
 we can write
\begin{eqnarray}
H_{0}^{(s)} &=&  v_{F} \int^{\infty}_{-\infty} \frac{dp}{2\pi } \,
|p|  \phi_{s}^{\dagger}(p) \phi_{s}(p)  ,       \\
\partial_{x}\Phi_{s}(x) &=&  i \int^{\infty}_{0} dp \sqrt{\frac{p}{2\pi}}
e^{-\alpha p /2}   \left[
\phi_{s}(p) e^{ipx} - \phi_{s}^{\dagger}(p) e^{-ipx} \right]  .
\end{eqnarray}
The commutation relation for the Fourier components is
\begin{equation}
[\phi_{s}(p), \phi_{s}^{\dagger}(p')] = 2\pi \delta(p-p') .  \label{crpp}
\end{equation}
 Next, let's introduce a generalized  transformation operator
\begin{equation}
\hat{U}(\lambda) = e^{-i\lambda S^{z}_{+} \Phi_{s}(0) }
= e^{-i \lambda S^{z}_{+} \int^{\infty}_{0}
dp e^{-\alpha p/2}
[\phi_{s}(p) + \phi_{s}^{\dagger}(p) ] /\sqrt{2\pi p} } ,
\end{equation}
and define two $\lambda$-dependent functions,
\begin{eqnarray}
f_{1}(\lambda) &=&  \hat{U}(\lambda)  H_{0}^{(s)}
				\hat{U}^{-1}(\lambda)  ,  \\
f_{2}(\lambda) &=&
	\hat{U}(\lambda)  \partial_{x} \Phi_{s}(x)
		\hat{U}^{-1}(\lambda) .
 \end{eqnarray}
 We note  $\hat{U}(\lambda=1)=\hat{U}$.
Using the commutation relation~(\ref{crpp}),
it is straightforward to verify
\begin{eqnarray}
\frac{d^{2}}{d\lambda^{2}} f_{1}(\lambda) &=&
\frac{2 v_{F}}{\alpha}  ( S^{z}_{+} )^{2} ,    \label{f1_1}  \\
\frac{d}{d\lambda} f_{1}(\lambda) |_{\lambda=0}  &=&
- v_{F} \partial_{x} \Phi_{s}(0) S^{z}_{+} ,   \label{f1_2}  \\
\frac{d}{d\lambda} f_{2}(\lambda) &=&
- 2 \int^{\infty}_{0} dp e^{-\alpha p} \cos(px) S^{z}_{+}  .  \label{f2_1}
\end{eqnarray}
{}From (\ref{f1_1}) and (\ref{f1_2}), we obtain
\begin{equation}
\hat{U}  H_{0}^{(s)} \hat{U}^{-1}
    = H_{0}^{(s)} - v_{F} \partial_{x} \Phi_{s}(0) S^{z}_{+}
    + \frac{v_{F}}{\alpha} ( S^{z}_{+})^{2} .     \label{tru1}
    \end{equation}
{}From (\ref{f2_1}), we obtain
\begin{equation}
	\hat{U}  \partial_{x} \Phi_{s}(x)
		\hat{U}^{-1}
   = \partial_{x} \Phi_{s}(x) - 2 \int^{\infty}_{0} dp e^{-\alpha p}
   \cos(px) S^{z}_{+} .
   \end{equation}
This implies
\begin{eqnarray}
	\hat{U}  \partial_{x} \Phi_{s}(0) \hat{U}^{-1}
  & =&  \partial_{x} \Phi_{s}(0) - \frac{2}{\alpha} S^{z}_{+} ,
			\label{tru2}    \\
   \hat{U}
   \int^{\infty}_{-\infty} dx \partial_{x} \Phi_{s}(x) \hat{U}^{-1}
 &=&   \int^{\infty}_{-\infty} dx \partial_{x} \Phi_{s}(x)
       - 2\pi S^{z}_{+}  .              \label{tru3}
       \end{eqnarray}
 Substituting (\ref{tru1}), (\ref{tru2}) and (\ref{tru3}) into
 $\hat{U} H \hat{U}^{-1} $, we obtain the results~(\ref{UH0U})
 and (\ref{H1_2}).

 \section{Derivation of the effective Hamiltonian in the
 second order projection }     \label{app_proj}

 In this appendix, we shall derive the
 coefficients $g_{0}$, $g_{1}$, $\widetilde{V}$,
$\alpha_{u}$ and $\alpha_{s}$
 in the effective Hamiltonian~(\ref{heff}),
 from the second order projection~(\ref{proj2}).
 The Hamiltonian to be projected is
 (\ref{h0_3}) + (\ref{h1_3}) + (\ref{h2_3}).

  The  first order contributions to the effective Hamiltonian are
  \begin{eqnarray}
  \hat{Q} H \hat{Q}
  &=& - i v_{F} \int^{\infty}_{-\infty} dx \sum_{\lambda=s,f,sf}
  \psi_{\lambda}^{\dagger}(x) \partial_{x} \psi_{\lambda}(x)
  + i \frac{v_{F} J_{m}^{\perp} }{\sqrt{2\pi\alpha} }
  \left[ \psi_{sf}(0)-\psi_{sf}^{\dagger}(0) \right]
  \hat{Q} S_{-}^{y} \hat{Q}
	\nonumber       \\
  & + & h_{u} \int^{\infty}_{-\infty}
 dx \psi_{s}^{\dagger}(x) \psi_{s}(x)  + H_{2} .
		\label{hproj1}
  \end{eqnarray}
  In terms of the local fermion operators $d$ and $d^{\dagger}$,
  we have $  \hat{Q} S_{-}^{y} \hat{Q}
  = i ( d - d^{\dagger} ) $.
  Performing the transformation~(\ref{ch_sta}) to
  install the anticommutation relations
  between the extended and local
  fermion operators,
  the hybridization term between the extended and local
  fermion operators in (\ref{hproj1})  becomes,
  \begin{equation}
  - \frac{v_{F} J_{m}^{\perp} }{\sqrt{2\pi\alpha} }
  \left[ \widetilde{\psi}_{sf}(0) -
  \widetilde{\psi}_{sf}^{\dagger}(0) \right] (d + d^{\dagger} ) .
			\label{heff1}
  \end{equation}

  In the second order of the projection~(\ref{proj2}),
  the local impurity spin state
  is virtually excited from one of the doublet to
  either $(|\uparrow\uparrow> - |\downarrow\downarrow>) /\sqrt{2}$
  or   $(|\uparrow\downarrow> + |\downarrow\uparrow>) /\sqrt{2}$,
  then returns back to the doublet.
  From Figure~\ref{loc_spin}, the mixing terms between the doublet
  and the excited states are those in (\ref{h0_3}) + (\ref{h1_3})
  which contain   $S_{+}^{x}$
  and $S_{\pm}^{z}$.
  If the local impurity spins leave and return
  to the same
  state of the doublet,
  the generated contributions to the effective Hamiltonian
  have the form of either $d^{\dagger}d$ or $d d^{\dagger}$.
  These are simply the  renormalizations to the RKKY interaction,
  shifting its critical value determined  by
  \begin{equation}
  - \widetilde{K}_{z} = K_{\perp} + \cdots ,
  \end{equation}
  where the omitted extra terms stand for the above mentioned
 renormalizations.
 The relevant term in the effective Hamiltonian
 around the critical point is,
 like the mass term in the usual critical phenomenon,
 \begin{equation}
 - \left( \frac{K_{z} + K_{\perp} }{2} - K_{c} \right) d^{\dagger} d,
  \hspace{.2in}    {\rm with} \; \;
  K_{c} = \frac{ v_{F} }{ \pi \alpha} \left( J_{+}^{z} - \pi \right)
	+ \cdots  .
 \end{equation}

  The other contributions come from the situations
  when the local impurity spins start from one state
  but return
  to the other state of the doublet.
  From Figure~\ref{loc_spin}, we see that these contributions
  must come from the projection of the product of $S_{+}^{x}$
  and $S_{-}^{z}$,
  \begin{eqnarray}
  & \hat{Q} &
  \left\{ \frac{v_{F} J^{z}_{m} }{2}
  \left[ \psi^{\dagger}_{sf}(0) \psi_{sf}(0)
   - \psi_{sf}(0) \psi_{sf}^{\dagger}(0)
  \right]
  S_{-}^{z}
  \right\}
			\nonumber               \\
  & & \times
   \frac{ 1-\hat{Q} } { -(K_{\perp} + T_{K} )
   + i v_{F} \int^{\infty}_{-\infty} dx \psi^{\dagger}_{sf}(x)
   \partial_{x} \psi_{sf}(x)
   - V \left[ \psi_{sf}(0) - \psi^{\dagger}_{sf}(0) \right]
   \left[ \psi_{f}(0) + \psi^{\dagger}_{f}(0) \right]     }
		\nonumber       \\
  & & \times
   \left\{ \frac{v_{F} J_{+}^{\perp} }{\sqrt{2\pi \alpha}}
   \left[ \psi_{sf}(0) + \psi^{\dagger}_{sf}(0) \right]
   S^{x}_{+} \right\}
  \hat{Q}
  +      \hat{Q}
  \left\{ \frac{v_{F} J_{+}^{\perp} }{\sqrt{2\pi \alpha}}
   \left[ \psi_{sf}(0) + \psi^{\dagger}_{sf}(0) \right]
   S^{x}_{+} \right\}
			\nonumber               \\
  & & \times
   \frac{ 1-\hat{Q} } { -(K_{\perp} + T_{K} )
   + i v_{F} \int^{\infty}_{-\infty} dx \psi^{\dagger}_{sf}(x)
   \partial_{x} \psi_{sf}(x)
   - V \left[ \psi_{sf}(0) - \psi^{\dagger}_{sf}(0) \right]
   \left[ \psi_{f}(0) + \psi^{\dagger}_{f}(0) \right]     }
		      \nonumber               \\
 & & \times
  \left\{ \frac{ v_{F} J^{z}_{m} }{2}
  \left[ \psi^{\dagger}_{sf}(0) \psi_{sf}(0)
   - \psi_{sf}(0) \psi_{sf}^{\dagger}(0)
  \right]
  S_{-}^{z}
  \right\} \hat{Q} .
		\label{sxsz}
   \end{eqnarray}
 Besides the energy gap $K_{\perp}+T_{K}$ between the doublet and
 the local  excited  state
 $(|\uparrow\uparrow> + |\downarrow\downarrow>) /\sqrt{2}$,
 we have also kept the intraband terms in the intermediate
 denominator which will be expanded as
 \begin{eqnarray}
    &  &   \left\{ -(K_{\perp} + T_{K} )
   + i v_{F} \int^{\infty}_{-\infty} dx \psi^{\dagger}_{sf}(x)
   \partial_{x} \psi_{sf}(x)
   - V \left[ \psi_{sf}(0) - \psi^{\dagger}_{sf}(0) \right]
   \left[ \psi_{f}(0) + \psi^{\dagger}_{f}(0) \right]     \right\}^{-1}
		  \nonumber               \\
 &  \simeq &
   - \frac{1}{K_{\perp}+T_{K}}
  - \frac{i v_{F} }{ (K_{\perp}+T_{K})^{2} }
    \int^{\infty}_{-\infty} dx \psi^{\dagger}_{sf}(x)
   \partial_{x} \psi_{sf}(x)
			\nonumber               \\
 & &   + \frac{ V }{ (K_{\perp}+T_{K})^{2} }
    \left[ \psi_{sf}(0) - \psi^{\dagger}_{sf}(0) \right]
   \left[ \psi_{f}(0) + \psi^{\dagger}_{f}(0) \right]  .
		\label{exp_med}
   \end{eqnarray}
Substituting (\ref{exp_med}) into (\ref{sxsz}),
we obtain  three contributions to the effective Hamiltonian.

The first contribution is
\begin{eqnarray}
 & &  -\frac{v_{F}^{2} J_{+}^{\perp}
 J_{m}^{z} }{2 (K_{\perp}+T_{K}) \sqrt{2\pi \alpha}}
 \left\{
 \left[ \psi_{sf}(0) + \psi^{\dagger}_{sf}(0) \right]
  \left[ \psi^{\dagger}_{sf}(0) \psi_{sf}(0)
   - \psi_{sf}(0) \psi_{sf}^{\dagger}(0)
  \right]
 \hat{Q} S_{+}^{x}    S_{-}^{z}  \hat{Q}
	\right.        \nonumber          \\
 & & \hspace{.1in}   \left.   +
  \left[ \psi^{\dagger}_{sf}(0) \psi_{sf}(0)
   - \psi_{sf}(0) \psi_{sf}^{\dagger}(0)
  \right]
  \left[ \psi_{sf}(0) + \psi^{\dagger}_{sf}(0) \right]
  \hat{Q} S_{-}^{z}    S_{+}^{x}  \hat{Q}
   \right\}  .                 \label{exp_med1}
 \end{eqnarray}
Using the fact
$  \hat{Q} S_{-}^{z}    S_{+}^{x}  \hat{Q}
  = d^{\dagger} $
  and
$  \hat{Q} S_{+}^{x}    S_{-}^{z}  \hat{Q}
  = d $,
 and  carrying out the transformation~(\ref{ch_sta}),
  we  simplify (\ref{exp_med1}) to
\begin{equation}
  - \frac{v_{F}^{2} J_{+}^{\perp}
  J_{m}^{z} }{2 (K_{\perp}+T_{K}) (2\pi \alpha)^{3/2} }
 \left[ \widetilde{\psi}_{sf}(0)
 - \widetilde{\psi}^{\dagger}_{sf}(0) \right] (d + d^{\dagger}) .
		\label{heff2}
 \end{equation}
In deriving (\ref{heff2}), we have  used the relation
\begin{equation}
\left\{   \psi_{sf}(0),  \psi^{\dagger}_{sf}(0) \right\}
= \delta(0) = \frac{1}{2\pi \alpha} .
\end{equation}
Noting that (\ref{heff2}) is a renormalization to (\ref{heff1}).

The second contribution to the effective Hamiltonian from
(\ref{sxsz}) is
\begin{eqnarray}
 &-& i  \frac{ v_{F}^{3} J_{+}^{\perp} J_{m}^{z}  }
 { 2 (K_{\perp}+T_{K})^{2} \sqrt{2\pi\alpha} }
 \left\{
  \left[ \psi_{sf}(0) + \psi^{\dagger}_{sf}(0) \right]
    \int^{\infty}_{-\infty} dx \psi^{\dagger}_{sf}(x)
   \partial_{x} \psi_{sf}(x)
	\right.        \nonumber          \\
 & & \times
 \left[ \psi^{\dagger}_{sf}(0) \psi_{sf}(0)
   - \psi_{sf}(0) \psi_{sf}^{\dagger}(0)
  \right]
 \hat{Q} S_{+}^{x}    S_{-}^{z}  \hat{Q}
  +   \left[ \psi^{\dagger}_{sf}(0) \psi_{sf}(0)
   - \psi_{sf}(0) \psi_{sf}^{\dagger}(0)
  \right]
	       \nonumber          \\
 & & \times     \left.
    \int^{\infty}_{-\infty} dx \psi^{\dagger}_{sf}(x)
   \partial_{x} \psi_{sf}(x)
   \left[ \psi_{sf}(0) + \psi^{\dagger}_{sf}(0) \right]
  \hat{Q} S_{-}^{z}    S_{+}^{x}  \hat{Q} \right\}.
			\label{exp_med2}
\end{eqnarray}
Commuting all fermion operators evaluated at $x=0$
to one side  and  simplifying the products using
anticommutation relations, we find that (\ref{exp_med2})
contains a term
\begin{equation}
 \frac{ v_{F}^{3} J_{+}^{\perp} J_{m}^{z} }
{ 8 (K_{\perp}+T_{K} )^{2}
(2\pi\alpha)^{3/2} } i \partial_{x}
   \left[ \widetilde{\psi}_{sf}(0) -
	\widetilde{\psi}^{\dagger}_{sf}(0) \right]
   (d - d^{\dagger})  .        \label{heff3}
 \end{equation}
 This is the leading irrelevant operator and has dimension 3/2.
 Noting that the combination of local fermion operators
 appearing in    (\ref{heff3}) is $d-d^{\dagger}$,
 not $d+d^{\dagger}$! This  is a vital difference.

The third contribution to the effective Hamiltonian from
(\ref{sxsz}) is
 \begin{eqnarray}
 & &   \frac{ V v_{F}^{2} J_{+}^{\perp} J_{m}^{z}  }
 { 2 (K_{\perp}+T_{K})^{2} \sqrt{2\pi\alpha} }
 \left\{
  \left[ \psi_{sf}(0) + \psi^{\dagger}_{sf}(0) \right]
   \left[ \psi_{sf}(0) - \psi^{\dagger}_{sf}(0) \right]
   \left[ \psi_{f}(0) + \psi^{\dagger}_{f}(0) \right]
	     \right.         \nonumber               \\
 & & \times
   \left[ \psi^{\dagger}_{sf}(0) \psi_{sf}(0)
   - \psi_{sf}(0) \psi_{sf}^{\dagger}(0)
  \right]
  \hat{Q} S_{+}^{x}    S_{-}^{z}  \hat{Q}
 + \left[ \psi^{\dagger}_{sf}(0) \psi_{sf}(0)
   - \psi_{sf}(0) \psi_{sf}^{\dagger}(0)
  \right]
		\nonumber               \\
 & & \times     \left.
   \left[ \psi_{sf}(0) - \psi^{\dagger}_{sf}(0) \right]
   \left[ \psi_{f}(0) + \psi^{\dagger}_{f}(0) \right]
  \left[ \psi_{sf}(0) + \psi^{\dagger}_{sf}(0) \right]
    \hat{Q} S_{+}^{x}    S_{-}^{z}  \hat{Q}
	     \right\} .
\end{eqnarray}
This contribution can be reduced to
\begin{equation}
 \frac{ V  v_{F}^{2} J_{+}^{\perp} J_{m}^{z} }
{ 2 (K_{\perp}+T_{K} )^{2}
(2\pi\alpha)^{5/2} }
   \left[ \widetilde{\psi}_{f}(0) +
	\widetilde{\psi}^{\dagger}_{f}(0) \right]
   (d - d^{\dagger})  .        \label{heff4}
 \end{equation}
Again,
 we note that it is $d - d^{\dagger}$ appearing in (\ref{heff4})!
 This is the second relevant operator which is present
 only when the particle-hole symmetry is broken,
 {\it i.e.} when $V \neq 0$.

The staggered magnetic field coupling term comes from
\begin{eqnarray}
& \hat{Q} &
\left\{ h_{s} S_{-}^{z}  \right\}
\frac{ 1 -\hat{Q} }{ -(K_{\perp}+T_{K} ) }
\left\{ \frac{v_{F} J_{+}^{\perp} }{\sqrt{2\pi\alpha}}
  \left[ \psi_{sf}(0) + \psi^{\dagger}_{sf}(0) \right]
 S_{+}^{x} \right\}      \hat{Q}
	\nonumber               \\
 &+&
 \hat{Q}    \left\{
\frac{v_{F} J_{+}^{\perp} }{\sqrt{2\pi\alpha}}
  \left[ \psi_{sf}(0) + \psi^{\dagger}_{sf}(0) \right]  \right\}
 \frac{ 1 -\hat{Q} }{ -(K_{\perp}+T_{K} ) }
\left\{  h_{s} S_{-}^{z}   \right\}
 \hat{Q}  .
 \end{eqnarray}
 This term is simplified to
 \begin{equation}
 - \frac{ h_{s} v_{F} J_{+}^{\perp} }
 { (K_{\perp}+T_{K}) \sqrt{2\pi\alpha} }
   \left[ \widetilde{\psi}_{sf}(0) +
	\widetilde{\psi}^{\dagger}_{sf}(0) \right]
   (d - d^{\dagger})  .        \label{heff5}
 \end{equation}

To obtain the uniform magnetic field coupling term,
we restore the bose field $\phi_{s}(x)$ through
$\psi_{s}^{\dagger}(x) \psi_{s}(x)
= \partial_{x} \Phi_{s}(x)/(2\pi)$
in  (\ref{h0_3}) and (\ref{h1_3}).
Then we integrate out $\phi_{s}(x)$ exactly.
This is carried out as follows.  First, we notice that
the terms in  (\ref{UH0U}) and (\ref{H1_2}) containing
$\phi_{s}$ can be rewritten as, upon inserting
$\partial_{x}\Phi_{s}(x) = \sqrt{\pi}
\left[ \partial_{x} \phi_{s}(x) - \Pi_{s}(x) \right]$,
\begin{eqnarray}
&  & \frac{v_{F}}{2}  \int^{\infty}_{-\infty} dx
\left\{ \Pi_{s}^{2}(x) + \left[ \partial_{x} \phi_{s}(x) \right]^{2}
+ \frac{h_{u}}{\pi v_{F} }   \partial_{x} \Phi_{s}(x)
+ \frac{\widetilde{J}_{+}^{z} }{\pi} \delta(x)
\partial_{x} \Phi_{s}(x) S_{+}^{z}
\right\}                 \nonumber              \\
 & =&  \frac{v_{F}}{2}  \int^{\infty}_{-\infty} dx
\left\{ \left[ \Pi_{s}(x)
- \frac{h_{u}}{2 \sqrt{\pi} v_{F}} \right]^{2}
+ \left[ \partial_{x} \phi_{s}(x)
  + \frac{h_{u}}{2 \sqrt{\pi} v_{F}} \right]^{2}
+ \frac{\widetilde{J}_{+}^{z} }{\pi} \delta(x)
\partial_{x} \Phi_{s}(x) S_{+}^{z}
\right\} ,
\label{med11}
\end{eqnarray}
up to an additive constant.
By introducing
\begin{equation}
\widetilde{\phi}_{s}(x) = \phi_{s}(x)
	+ \frac{h_{u}}{2 \sqrt{\pi} v_{F}}  x  ,
\end{equation}
we can recast (\ref{med11}) in the form
\begin{equation}
\frac{v_{F}}{2} \int^{\infty}_{-\infty} dx
\left\{ \widetilde{\Pi}_{s}^{2}(x) +
\left[ \partial_{x} \widetilde{\phi}_{s}(x) \right]^{2}
+ \frac{\widetilde{J}_{+}^{z} }{\pi} \delta(x)
\partial_{x} \widetilde{\Phi}_{s}(x) S_{+}^{z}
- \frac{h_{u} \widetilde{J}_{+}^{z} }{\pi v_{F} } S_{+}^{z} \delta(x)
\right\}  ,                \label{med12}
\end{equation}
where $ \widetilde{\Pi}_{s}(x) $ and
$\partial_{x} \widetilde{\Phi}_{s}(x)$ are correspondingly
defined as, in consistency with (\ref{def_Phi}) and the
relation $\widetilde{\Pi}_{s} = \partial_{t} \widetilde{\phi}_{s}$,
\begin{eqnarray}
    \widetilde{\Pi}_{s}(x) &=& \Pi_{s}(x) -
		\frac{h_{u}}{2 \sqrt{\pi} v_{F}}   ,     \\
   \partial_{x} \widetilde{\Phi}_{s}(x)  &=& \partial_{x} \Phi_{s}(x)
   +\frac{h_{u}}{v_{F}} .
   \end{eqnarray}
 Since the uniform field $h_{u}$ only appears in
the last term of  (\ref{med12}),
we only need to project it onto the lowest doublet in the next step.
The contribution is
\begin{eqnarray}
&\hat{Q} &
\left\{
\frac{v_{F} J_{-}^{z} }{2}
   \left[ \psi_{sf}(0) + \psi^{\dagger}_{sf}(0) \right]
   \left[ \psi_{f}(0) - \psi^{\dagger}_{f}(0) \right]
   S_{+}^{z}
 + v_{F} \widetilde{J}_{+}^{z} \psi^{\dagger}_{s}(0) \psi_{s}(0) S_{+}^{z}
\right\}
		\nonumber       \\
& \times &
\frac{ 1-\hat{Q} }{ - T_{K} }
\left\{ -
\frac{ h_{u} \widetilde{J}_{+}^{z} }{ 2 \pi}
S_{+}^{z}
\right\} \hat{Q}
+ \hat{Q}  \left\{ -
\frac{ h_{u} \widetilde{J}_{+}^{z} }{ 2 \pi}
S_{+}^{z}
\right\}
\frac{ 1-\hat{Q} }{ - T_{K} }
		\nonumber       \\
& \times &
\left\{
\frac{v_{F} J_{-}^{z} }{2}
   \left[ \psi_{sf}(0) + \psi^{\dagger}_{sf}(0) \right]
   \left[ \psi_{f}(0) - \psi^{\dagger}_{f}(0) \right]
   S_{+}^{z}
 + v_{F} \widetilde{J}_{+}^{z} \psi^{\dagger}_{s}(0) \psi_{s}(0) S_{+}^{z}
\right\} \hat{Q}  .         \label{med13}
\end{eqnarray}
With a little algebra, one can  show that (\ref{med13})
contains
\begin{equation}
\frac{ h_{u} v_{F} \widetilde{J}_{+}^{z} J_{-}^{z} }{ 8 \pi T_{K} }
   \left[ \psi_{sf}(0) + \psi^{\dagger}_{sf}(0) \right]
   \left[ \psi_{f}(0) - \psi^{\dagger}_{f}(0) \right]
+ \frac{h_{u} v_{F}  (\widetilde{J}_{+}^{z})^{2}  }{4\pi T_{K} }
\psi^{\dagger}_{s}(0) \psi_{s}(0) .
   \label{heff6}
\end{equation}

Combining the results (\ref{heff1}),  (\ref{heff2}), (\ref{heff3}),
(\ref{heff4}), (\ref{heff5}), (\ref{heff6}) together
and omitting the tilde signs on $\psi$'s,
we obtain the effective Hamiltonian~(\ref{HFP}), (\ref{HPERT})
and (\ref{hphb}).
The qualitative physics does not depend on
the numerical values of the coefficients
$g_{0}$, $g_{1}$,  $\widetilde{V}$,
$\alpha_{u}$ and  $\alpha_{s}$
in the effective Hamiltonian.
It should be kept in mind that
the obtained expressions,  (\ref{def_g0}) to (\ref{def_alp_s}),
for the numerical coefficients of the
effective Hamiltonian
 should not be taken too literally.
The purpose of this appendix is to illustrate how each
term in the effective Hamiltonian arises from the
projection rather than accurately determining the coefficients
of the effective Hamiltonian.
A practical way to determine them  probably is to
fit numerical results or experimental data.

 \section{ Effect of the marginal operators
(\ref{HPHB_MORE})  }      \label{more_phb}

In this appendix, we shall show that the only
effect of including the marginal particle-hole
symmetry breaking operators (\ref{HPHB_MORE}) is to slightly
renormalize the two basics energy scales $T_{K}$ and
$T_{c}$.

The marginal operators~(\ref{HPHB_MORE}) correspond to
 the following  terms in the action,
\begin{equation}
{\cal S}'_{phb}= 2iV \sum_{n}
\int^{\infty}_{-\infty}
\frac{dk}{2\pi} a_{f}(\omega_{n},k) \left[
\alpha_{v} a(-\omega_{n}) -   \int^{\infty}_{-\infty}
\frac{dk'}{2\pi} b_{sf}(-\omega_{n},k') \right] .
\end{equation}
Our task now is to diagonalize
\begin{equation}
{\cal S}(b_{sf},a_{f}) = {\cal S}_{2}(b_{sf}) + {\cal S}_{3}(a_{f})
+ {\cal S}'_{phb}  ,            \label{sbsfaf}
\end{equation}
where ${\cal S}_{2}(b_{sf})$ and $ {\cal S}_{3}(a_{f})$ are given
by (\ref{S2}) and (\ref{S3}) respectively.
More specifically, we need to find new linear transformations
other than (\ref{shift_bsf}) and (\ref{shift_af}) such that
the hybridizing terms in (\ref{sbsfaf}) are canceled out.
The desired transformations are
\begin{eqnarray}
 \widetilde{b}_{sf}(\omega_{n},k) &=&    b_{sf}(\omega_{n},k)
 + \xi_{1}(\omega_{n},k) \, a(\omega_{n})
 + \xi_{2}(\omega_{n},k) \, b(\omega_{n})    ,          \\
 \widetilde{a}_{f}(\omega_{n},k) &=&    a_{f}(\omega_{n},k)
 + \xi_{3}(\omega_{n},k) \, a(\omega_{n})
 + \xi_{4}(\omega_{n},k) \, b(\omega_{n})    ,
 \end{eqnarray}
 where the four transformation coefficients are given by
 \begin{eqnarray}
 \xi_{1}(\omega_{n},k) &=& - \frac{2iv_{F} }{ i\omega_{n} - v_{F} k}
 \frac{ g_{0} + {\rm sgn}\omega_{n} \,
 \alpha_{v} V^{2}/v_{F}^{2}  }{
    1+ V^{2}/v_{F}^{2} }   ,            \\
 \xi_{2}(\omega_{n},k) &=& \frac{2v_{F} }{ i\omega_{n} - v_{F} k}
 \left[ g_{1} k - \frac{i V {\rm sgn}\omega_{n} }{ v_{F}^{2} + V^{2} }
     \left( \widetilde{V} - \frac{2V g_{1} \Lambda}{\pi}
     +\frac{Vg_{1}}{v_{F}} |\omega_{n}|
     \right) \right]  ,      \\
 \xi_{3}(\omega_{n},k) &=& - \frac{2i V }{ i\omega_{n} - v_{F} k}
 \frac{ \alpha_{v} -g_{0}\, {\rm sgn}\omega_{n} }{1+ V^{2}/v_{F}^{2}} ,  \\
 \xi_{4}(\omega_{n},k) &=& - \frac{2i }{ i\omega_{n} - v_{F} k}
   \frac{ \widetilde{V} - 2V g_{1} \Lambda/\pi
     + Vg_{1} |\omega_{n}|/v_{F}   }{ 1 + V^{2}/v_{F}^{2} }  .
 \end{eqnarray}
 The ultraviolet cutoff $\Lambda$ enters
the transformation coefficients through the
 integrals~(\ref{integ2}) and (\ref{integ3}).
In terms of the shifted Grassmann variables,
the action~(\ref{sbsfaf}) becomes
\begin{eqnarray}
  {\cal S}(b_{sf},a_{f}) &=&
  - \sum_{n} \int^{\infty}_{-\infty} \frac{d k}{2\pi}
\left\{ \frac{1}{2}
(i\omega_{n} - v_{F} k)
\left[  \widetilde{a}_{f}(-\omega_{n} , -k)
	\widetilde{a}_{f}(\omega_{n} , k)
     +  \widetilde{b}_{sf}(-\omega_{n} , -k)
	\widetilde{b}_{sf}(\omega_{n} , k)    \right]
	\right.        \nonumber               \\
 & & \left.
 + 2iV  \int^{\infty}_{-\infty} \frac{d k'}{2\pi}
   \widetilde{a}_{f}(\omega_{n} , k)
   \widetilde{b}_{sf}(-\omega_{n} , k')   \right\}
   + {\cal S}_{gen}(a,b)  .
\end{eqnarray}
The generated local terms are
\begin{eqnarray}
{\cal S}_{gen}(a,b) &=& - i \sum_{n}
\left\{ \frac{ {\rm sgn}\omega_{n} }{2} \left[
  {\cal M}_{aa}(|\omega_{n}|) a(-\omega_{n}) a(\omega_{n})
+ {\cal M}_{bb}(|\omega_{n}|) b(-\omega_{n}) b(\omega_{n})   \right]
		\right.         \nonumber               \\
& & \left.
+ {\cal M}_{ab}( \omega_{n} ) a(-\omega_{n}) b(\omega_{n})   \right\} ,
		\label{sgen}
\end{eqnarray}
with
\begin{eqnarray}
{\cal M}_{aa}(|\omega_{n}|) &=& \frac{ 2v_{F} }{ 1+V^{2}/v_{F}^{2}}
   \left[ g_{0}^{2} + \left( \frac{ \alpha_{v} V }{v_{F}} \right)^{2}
   \right] ,            \\
{\cal M}_{bb}(|\omega_{n}|) &=& \frac{ 2 }{ 1+V^{2}/v_{F}^{2} }
   \left\{ \frac{1}{v_{F}} \left( \widetilde{V}
   - \frac{2V g_{1} \Lambda}{\pi} \right)^{2}
		\right.         \nonumber       \\
& & \left.
   + |\omega_{n}| \left[ \frac{2g_{1}^{2} \Lambda}{\pi}
   \left(1-\frac{V^{2}}{v_{F}^{2}}\right)
   + \frac{ 2g_{1} V \widetilde{V} }{ v_{F}^{2} } \right]
   - \frac{g_{1}^{2}}{v_{F}} \omega_{n}^{2}  \right\}  ,        \\
{\cal M}_{ab}( \omega_{n} ) &=& \frac{2}{1 + V^{2}/v_{F}^{2} }
   \left[ g_{0} \left( \frac{V\widetilde{V}}{v_{F}}
  + \frac{2v_{F}g_{1} \Lambda }{\pi} - g_{1} |\omega_{n}| \right)
		\right.         \nonumber       \\
& & \left.
  + {\rm sgn}\omega_{n} \frac{\alpha_{v} V}{v_{F}}  \left(
  \widetilde{V} - \frac{2V g_{1} \Lambda }{\pi}
  + \frac{ V g_{1}}{v_{F}} |\omega_{n}|  \right)  \right]  .
\end{eqnarray}
To  obtain the total effective local action, we
add the generated terms~(\ref{sgen})  to (\ref{slocal}).
Introducing a new local Grassmann variable,
\begin{equation}
\widetilde{b}(\omega_{n}) = b(\omega_{n}) + \frac{\alpha_{v} V }{
\widetilde{V} - 2V g_{1} \Lambda /\pi } a(\omega_{n}) ,
\end{equation}
we can compactly write the effective local action as,
upon neglecting $\omega_{n}^{2}$ terms in the matrix elements,
\begin{eqnarray}
{\cal S}_{loc}^{eff} &=&
 - i \, \sum_{n>0}
 \left(  a( -\omega_{n}),  \widetilde{b}( -\omega_{n}) \right)
		\nonumber       \\
 & \times &
  \left( \begin{array}{cc}
 \omega_{n} /Z_{a} + 2v_{F} \widetilde{g}_{0}^{2}   &
- \left( \widetilde{\delta K}
+ 2 \widetilde{g}_{0} \widetilde{g}_{1} \omega_{n}   \right)
     - \eta \omega_{n}       \\
 \widetilde{\delta K}
+ 2 \widetilde{g}_{0} \widetilde{g}_{1} \omega_{n}
- \eta \omega_{n}       &
   \omega_{n} / Z_{b}
+ 2  \widetilde{V'}^{2} /v_{F}      \\
\end{array}     \right)
\left(  \begin{array}{c}
	 a( \omega_{n}) \\   \widetilde{b}( \omega_{n})
	\end{array}     \right)  .      \label{seff_loc1}
\end{eqnarray}
The renormalized parameters are,
\begin{eqnarray}
\widetilde{g}_{0} &=& \frac{g_{0} }{ \sqrt{1+ V^{2}/v_{F}^{2}} } ,  \\
\widetilde{g}_{1} &=& \frac{g_{1} }{ \sqrt{1+ V^{2}/v_{F}^{2}} } ,  \\
\widetilde{V'} &=& \frac{ \widetilde{V} - 2Vg_{1}\Lambda/\pi }
	  { \sqrt{1+ V^{2}/v_{F}^{2}} } ,               \\
\widetilde{\delta K} &=& \delta K - \frac{2v_{F}g_{0}}{1+V^{2}/v_{F}^{2}}
\left(\frac{V\widetilde{V}}{v_{F}^{2}} + \frac{2g_{1}\Lambda}{\pi} \right) ,
\\
\frac{1}{Z_{a}} &=& 1+
\left( \frac{\alpha_{v}V}{\widetilde{V}-2Vg_{1}\Lambda/\pi} \right)^{2}
\left(1+\frac{4g_{1}^{2}\Lambda }{\pi} \right) ,                \\
\frac{1}{Z_{b}} &=& 1+ \frac{2}{1+V^{2}/v_{F}^{2} }
\left[ \frac{2g_{1}^{2} \Lambda }{\pi}
\left(1-\frac{V^{2}}{v_{F}^{2}}\right)
+ \frac{ 2g_{1} V\widetilde{V}}{v_{F}^{2}} \right] ,              \\
\eta &=&  \frac{\alpha_{v}V}{\widetilde{V}-2Vg_{1}\Lambda/\pi}
  \left[ 1 +
  \frac{ 2( 2g_{1}^{2}\Lambda/\pi + g_{1} V \widetilde{V}/v_{F}^{2} ) }
       {1+V^{2}/v_{F}^{2} }     \right] .
 \end{eqnarray}
The effective action~(\ref{seff_loc1}) has essentially the same form
as (\ref{seff_loc}) except for
a wave function renormalization factor $Z_{a}$ and
a new type term, $\eta \omega_{n}$, in
the off-diagonal matrix elements. However, this new type term
is irrelevant since it can only generate a $(\eta \omega_{n})^{2}$
term in physical quantities such as free energy.
At this point, it is clear that including the marginal particle-hole
symmetry breaking terms with the coefficients $V$ and $\alpha_{v}V$
will only renormalize the two  energy scales, the Kondo temperature
$T_{K}$ and the crossover temperature $T_{c}$.

\section{ Contribution of the marginal operators     (\ref{HSTAG_MORE})
to the staggered susceptibility}   \label{more_chis}

In this appendix, we show that  the contributions to the
staggered susceptibility from the marginal operators~(\ref{HSTAG_MORE})
are negligible.

The marginal operators~(\ref{HSTAG_MORE}) have the following
corresponding terms in the action
\begin{equation}
{\cal S}'_{stag} = 2i h_{s} \sum_{n}
 \int^{\infty}_{-\infty} \frac{d k}{2\pi}
a_{sf}(-\omega_{n},k) \left[ \alpha'_{s}
\int^{\infty}_{-\infty} \frac{d k'}{2\pi} b_{sf}(\omega_{n},k')
+\alpha''_{s} a(\omega_{n})     \right] .
\end{equation}
Combining the last expression
 with the staggered field coupling term in (\ref{S1})
and  inserting the transformation~(\ref{shift_bsf}),
we can write the complete staggered field coupling terms in the
following form
\begin{eqnarray}
{\cal S}_{stag} &=& 2i  h_{s}  \sum_{n}
 \int^{\infty}_{-\infty} \frac{d k}{2\pi}
a_{sf}(-\omega_{n},k)  \left[
\alpha'_{s}  \int^{\infty}_{-\infty} \frac{d k'}{2\pi}
 \widetilde{b}_{sf}(\omega_{n},k')
	\right.         \nonumber       \\
& &  \left.
+ \left( \alpha''_{s}
+ \alpha'_{s} g_{0}  \, {\rm sgn}\omega_{n} \right) a(\omega_{n})
+ \left( \alpha_{s} + \frac{2 \alpha'_{s} g_{1} \Lambda }{ \pi }
-\frac{ \alpha'_{s} g_{1} |\omega_{n}|}{ v_{F} }  \right)
	 b(\omega_{n}) \right] .        \label{s_stag_tot}
\end{eqnarray}
The staggered susceptibility is obtained by calculating
the second order perturbation of ${\cal S}_{stag}$,
\begin{equation}
\chi_{s} = - \frac{\partial^{2} }{\partial h_{s}^{2} }
\left[ - \frac{1}{2\beta}
\langle {\cal S}_{stag} {\cal S}_{stag} \rangle  \right] ,
			\label{chis_ful}
\end{equation}
where the average is
weighted by  an action consisting of the free  and decoupled
Grassmann variables $a_{sf}$, $\widetilde{b}_{sf}$
and the effective local action~(\ref{seff_loc}).
The first term in ${\cal S}_{stag}$ is a potential scattering term
and does not mix with the other terms of ${\cal S}_{stag}$
in the second order perturbation.
It thus gives
a finite contribution to the staggered susceptibility
and can be treated separately.
Carrying out the calculation for (\ref{chis_ful}), we find that
the singular part of the staggered susceptibility is
\begin{equation}
\chi_{s} = \frac{2}{v_{F}} \left( \alpha_{s} +
\frac{2\alpha'_{s}}{\pi} g_{1} \Lambda \right)^{2}
\frac{1}{\beta} \sum_{n}  i \, {\rm sgn}\omega_{n}
\langle b(-\omega_{n}) b(\omega_{n}) \rangle .  \label{chis_all}
\end{equation}
The propagator is given by, from (\ref{seff_loc}),
\begin{equation}
\langle b(-\omega_{n}) b(\omega_{n}) \rangle
= - i \, {\rm sgn}\omega_{n}
\frac{ Z_{b} \left( |\omega_{n}| + T_{K} \right) }
{ \omega_{n}^{2} + T_{c}^{2} + |\omega_{n}| T_{K} } , \label{green_bb}
\end{equation}
where we have taken $2v_{F} g_{0}^{2} \simeq T_{K}$
in the numerator of  (\ref{green_bb}) for simplicity.
The singularity of the staggered susceptibility comes from
the fact that at the critical point, $T_{c}=0$,
$\langle b(-\omega_{n}) b(\omega_{n}) \rangle \sim 1/(i\omega_{n}) $
which gives rise to the logarithmic singularity
for the  Matsubara frequency summation in (\ref{chis_all}).
{}From (\ref{chis_all}), we see that the only effect of the marginal
operators~(\ref{HSTAG_MORE}) is to shift $\alpha_{s}$
to $\alpha_{s}+2\alpha'_{s}g_{1} \Lambda/\pi$.
Actually, one should be able to see this
from (\ref{s_stag_tot}) without doing
calculation.


\begin{table}
\caption{Definition of frequently used parameters and symbols }
\begin{tabular}{cccc}
 Symbol  & Definition (Eq. No.) &  Symbol & Definition (Eq. No.)        \\
\tableline
 $v_{F}$                 &  Fermi velocity        &
		$ g_{1} $        & (\ref{HPERT}), (\ref{def_g1})        \\

 $\rho_{F}$    &  Density of states &
		$\widetilde{V} $   &  (\ref{hphb}), (\ref{def_tilde_v})  \\

 $J^{z}, \; J^{\perp} $ & Kondo coupling constants              &
		$ \alpha_{u} $  & (\ref{HFP}), (\ref{def_alp_u})         \\

  $ K_{z}, \; K_{\perp} $            &   RKKY interaction &
		$\alpha_{s}$     & (\ref{HFP}), (\ref{def_alp_s})        \\

 $S_{\pm}^{x,y,z}$  &   (\ref{spmxyz})  &
		$\delta K$       &  (\ref{def_dK})                      \\

  $h_{u}, \; h_{s} $  & (\ref{h01}) &
		$\Lambda $       &  (\ref{integ2})                       \\

 $ J_{\pm}^{z,\perp}, \; J_{m}^{z,\perp} $ & (\ref{coupl_const})  &
		$Z_{b}$  &  (\ref{def_zb})                               \\

 $ V $ & (\ref{ph-break})  &
		$T_{K}$    &    (\ref{def_tk})                           \\

 $ \widetilde{J}_{+}^{z}, \; \widetilde{K}_{z} $ &  (\ref{tildeK})   &
		$T_{c}$   &   (\ref{def_tc})                             \\

  $ g_{0} $   & (\ref{HFP}), (\ref{def_g0})     &
		$\widetilde{\alpha}_{s}$        & (\ref{def_tilde_as})  \\
\end{tabular}
\label{notations}
\end{table}

\clearpage

\begin{table}
\caption{The building blocks for constructing  operators
around the critical point.}
\begin{tabular}{ccccc}
 Operator  & Dimension &  Parity & Particle-hole & $\pi$ x-axis \\
\tableline
$\psi_{sf}(0)-\psi^{\dagger}_{sf}(0)$   & 1/2 & $-$ & +   & $-$ \\
$\psi_{sf}(0)+\psi^{\dagger}_{sf}(0)$   & 1/2 &  +  & +   & +   \\
$\psi_{f}(0)+\psi^{\dagger}_{f}(0)$     & 1/2 & $-$ & $-$ & $-$ \\
$\psi_{f}(0)-\psi^{\dagger}_{f}(0)$     & 1/2 &   + &  +  & $-$ \\
$d+d^{\dagger}$                         & 1/2 & $-$ & +   & $-$ \\
$d-d^{\dagger}$                         & 0   & $-$ & +   & $-$ \\
$  \psi_{s}^{\dagger}(0) \psi_{s}(0)
 - \psi_{s}(0) \psi_{s}^{\dagger}(0)$   & 1   & +   & +   & $-$ \\
$ \partial_{x} $                        & 1   & +   & +   & +   \\
\end{tabular}
\label{blocks}
\end{table}

\clearpage

\begin{table}
\caption{       All dimension 1/2 operators.
 The first one is the relevant operator. The second one
could couple to the staggered field. The third one could become
the second relevant operator
if the particle-hole symmetry is broken.
 As explained in the text,
$[\psi_{sf}(0)-\psi^{\dagger}_{sf}(0)](d-d^{\dagger})$
does not exist.
This is because    $[\psi_{sf}(0)-\psi^{\dagger}_{sf}(0)]$
could hybridize with either $d+d^{\dagger}$ or $d-d^{\dagger}$, but
not both. }
.
\begin{tabular}{cccc}
 Operator  &   Parity & Particle-hole & $\pi$ x-axis \\
\tableline
$(d+d^{\dagger})(d-d^{\dagger})$       &   +    &  +  & +       \\
$[\psi_{sf}(0)+\psi^{\dagger}_{sf}(0)](d-d^{\dagger})$
					&  $-$   & +   & $-$     \\
$[\psi_{f}(0)+\psi^{\dagger}_{f}(0)](d-d^{\dagger})$
					&   +    & $-$ & +       \\
$[\psi_{f}(0)-\psi^{\dagger}_{f}(0)](d-d^{\dagger})$
					&  $-$   &  +  & +       \\
\end{tabular}
\label{dim12}
\end{table}

\begin{table}
\caption{All dimension 1 operators. The first one is the hybridization term.
The second one can couple to the uniform magnetic field $h_{u}$.
The third and fourth operators could couple to the staggered field
$h_{s}$. The fifth and ninth are
the marginal particle-hole symmetry breaking operators.}
.
\begin{tabular}{cccc}
 Operator  &   Parity & Particle-hole & $\pi$ x-axis \\
\tableline
$[\psi_{sf}(0)-\psi^{\dagger}_{sf}(0)] (d+d^{\dagger}) $
					&   +    & +   &  +      \\
$[\psi_{sf}(0)+\psi^{\dagger}_{sf}(0)]
	[\psi_{f}(0)-\psi^{\dagger}_{f}(0)] $
					&   +    & +   & $-$     \\
$[\psi_{sf}(0)-\psi^{\dagger}_{sf}(0)]
      [\psi_{sf}(0)+\psi^{\dagger}_{sf}(0)] $
					&  $-$   & +   & $-$     \\
$[\psi_{sf}(0)+\psi^{\dagger}_{sf}(0)] (d+d^{\dagger}) $
					&  $-$   & +   & $-$     \\
$[\psi_{sf}(0)-\psi^{\dagger}_{sf}(0)]
	[\psi_{f}(0)+\psi^{\dagger}_{f}(0)] $
					&   +    & $-$ &  +      \\
$[\psi_{sf}(0)-\psi^{\dagger}_{sf}(0)]
	[\psi_{f}(0)-\psi^{\dagger}_{f}(0)] $
					&  $-$   & +   &  +      \\
$[\psi_{sf}(0)+\psi^{\dagger}_{sf}(0)]
	[\psi_{f}(0)+\psi^{\dagger}_{f}(0)] $
					&  $-$   & $-$ & $-$     \\
$[\psi_{f}(0)+\psi^{\dagger}_{f}(0)]
	[\psi_{f}(0)-\psi^{\dagger}_{f}(0)] $
					&  $-$   & $-$ &  +      \\
$[\psi_{f}(0)+\psi^{\dagger}_{f}(0)] (d+d^{\dagger}) $
					&   +    & $-$ &  +      \\

$[\psi_{f}(0)-\psi^{\dagger}_{f}(0)] (d+d^{\dagger}) $
					&  $-$   & +   &  +      \\
$  \psi_{s}^{\dagger}(0) \psi_{s}(0)
 - \psi_{s}(0) \psi_{s}^{\dagger}(0)$  & + & + & $-$  \\
\end{tabular}
\label{dim1}
\end{table}

\begin{table}
\caption{All dimension 3/2 operators. The first one  is the only
allowed leading irrelevant operator in the presence of
particle-hole symmetry. }
.
\begin{tabular}{cccc}
 Operator  &   Parity & Particle-hole & $\pi$ x-axis \\
\tableline

$ \partial_{x} [\psi_{sf}(0)-\psi^{\dagger}_{sf}(0)] (d-d^{\dagger}) $
					&   +    & +   &  +      \\                                         $
\partial_{x} [\psi_{sf}(0)+\psi^{\dagger}_{sf}(0)] (d-d^{\dagger}) $
					&  $-$   & +   & $-$      \\
 $ \partial_{x} [\psi_{f}(0)+\psi^{\dagger}_{f}(0)] (d-d^{\dagger}) $
					 &   +    &$-$  &  +      \\                                   $
\partial_{x} [\psi_{f}(0)-\psi^{\dagger}_{f}(0)] (d-d^{\dagger}) $
					&  $-$   & +   &  +      \\
$ [ \psi^{\dagger}_{s}(0)\psi_{s}(0)
- \psi_{s}(0)\psi^{\dagger}_{s}(0) ]
   [\psi_{sf}(0)-\psi^{\dagger}_{sf}(0)] (d-d^{\dagger}) $
				       &   +   & +   &  $-$      \\
$ [ \psi^{\dagger}_{s}(0)\psi_{s}(0)
- \psi_{s}(0)\psi^{\dagger}_{s}(0) ]
  [\psi_{sf}(0)+\psi^{\dagger}_{sf}(0)] (d-d^{\dagger}) $
					&  $-$   & +   &  +      \\
$ [ \psi^{\dagger}_{s}(0)\psi_{s}(0)
- \psi_{s}(0)\psi^{\dagger}_{s}(0) ]
  [\psi_{f}(0)+\psi^{\dagger}_{f}(0)] (d-d^{\dagger}) $
					&   +   &$-$  &  $-$     \\
$ [ \psi^{\dagger}_{s}(0)\psi_{s}(0)
- \psi_{s}(0)\psi^{\dagger}_{s}(0) ]
  [\psi_{f}(0)-\psi^{\dagger}_{f}(0)] (d-d^{\dagger}) $
					&  $-$  & +   & $-$      \\
$ [ \psi^{\dagger}_{s}(0)\psi_{s}(0)
- \psi_{s}(0)\psi^{\dagger}_{s}(0) ]
(d+d^{\dagger}) (d-d^{\dagger}) $
					&   +   & +   & $-$     \\
$  [\psi_{sf}(0)-\psi^{\dagger}_{sf}(0)]
	[\psi_{sf}(0)+\psi^{\dagger}_{sf}(0)]
	[\psi_{f}(0)+\psi^{\dagger}_{f}(0)]
 (d-d^{\dagger}) $
					&  $-$  &$-$  & $-$     \\
$  [\psi_{sf}(0)-\psi^{\dagger}_{sf}(0)]
	[\psi_{sf}(0)+\psi^{\dagger}_{sf}(0)]
	[\psi_{f}(0)-\psi^{\dagger}_{f}(0)]
 (d-d^{\dagger}) $
					&   +   & +  & $-$     \\
$  [\psi_{sf}(0)-\psi^{\dagger}_{sf}(0)]
	[\psi_{sf}(0)+\psi^{\dagger}_{sf}(0)] (d+d^{\dagger})
 (d-d^{\dagger}) $
					&  $-$  & +  & $-$     \\
$ [\psi_{sf}(0)+\psi^{\dagger}_{sf}(0)]
	[\psi_{f}(0)+\psi^{\dagger}_{f}(0)]
	[\psi_{f}(0)-\psi^{\dagger}_{f}(0)]
 (d-d^{\dagger})$
					&   +   &$-$  & $-$     \\
$ [\psi_{sf}(0)+\psi^{\dagger}_{sf}(0)]
	[\psi_{f}(0)+\psi^{\dagger}_{f}(0)] (d+d^{\dagger})
 (d-d^{\dagger}) $
					&  $-$  &$-$  & $-$     \\
$  [\psi_{f}(0)+\psi^{\dagger}_{f}(0)]
	[\psi_{f}(0)-\psi^{\dagger}_{f}(0)] (d+d^{\dagger})
 (d-d^{\dagger})$
					&  $-$  &$-$  &  +      \\
$  [\psi_{sf}(0)-\psi^{\dagger}_{sf}(0)]
	[\psi_{f}(0)+\psi^{\dagger}_{f}(0)]
	[\psi_{f}(0)-\psi^{\dagger}_{f}(0)]
 (d-d^{\dagger})$
					&  $-$  &$-$  &  +      \\
$ [\psi_{sf}(0)-\psi^{\dagger}_{sf}(0)]
	[\psi_{f}(0)+\psi^{\dagger}_{f}(0)] (d+d^{\dagger})
 (d-d^{\dagger}) $
					&   +   &$-$  &  +      \\
$ [\psi_{sf}(0)-\psi^{\dagger}_{sf}(0)]
	[\psi_{f}(0)-\psi^{\dagger}_{f}(0)] (d+d^{\dagger})
 (d-d^{\dagger}) $
					&  $-$  & +   &  +     \\
$ [\psi_{sf}(0)+\psi^{\dagger}_{sf}(0)]
	[\psi_{f}(0)-\psi^{\dagger}_{f}(0)] (d+d^{\dagger})
 (d-d^{\dagger}) $
					&   +   & +   & $-$     \\
\end{tabular}
\label{dim32}
\end{table}

\begin{figure}
\caption{
  The phase diagram of the two-impurity Kondo model.
  $V$ is the energy
  scale characterizing particle-hole symmetry breaking strength.
  $K$ is the fully
renormalized RKKY interaction. $T_K $ is the Kondo temperature.
  Except at the critical point marked by the black dot,
  the low energy behavior is Fermi-liquid type everywhere.
  The shaded area is the region where our solution applies.
The radius of
the solution region  is a fraction of $T_K $.   }
  \label{phase_diag}
  \end{figure}

\begin{figure}
\caption{
The energy level scheme of the four impurity spin states.
The up and down spin states
refer to the eigenstates of the operators $ S_1^z $
and  $ S_2^z $.
At $ - \widetilde{K}_z = K_\perp $, the two levels
   $ (|  \uparrow\downarrow  > - \, |  \downarrow\uparrow  >)
	/ \protect{\sqrt 2} $
	 and
   $ (|  \uparrow\uparrow  > + \, | \downarrow\downarrow  >)
	/ \protect{\sqrt 2} $
become degenerate, forming a doublet.
The
superficial degeneracy between the doublet and
	 $ ( |\uparrow\uparrow> - \, |\downarrow\downarrow> )
	/ \protect{\sqrt 2} $
is lifted by the
Kondo interaction term $ J_m^\perp $.       }
  \label{loc_lev}
  \end{figure}

\begin{figure}
\caption{
The four impurity-spin states and the
impurity spin operators connecting them,
$S_\pm^\lambda = S^\lambda_1  \pm
S^\lambda_2 $ for $\lambda=x,y,z$.      }
  \label{loc_spin}
  \end{figure}

\begin{figure}
\caption{
 The crossover function for the specific heat,
 Eq.\ (\protect{\ref{exp_ccross}}),
for various
values of $T_c / T_K $.  $ T_K $ is the Kondo temperature, and
$ T_c $ is the crossover temperature.   }
  \label{c_cross}
  \end{figure}

\begin{figure}
\caption{
 The crossover function for the entropy, Eq.\ (\protect{\ref{fe3}}),
for various
values of $ T_c /T_K $.     }
  \label{entrop_cross}
  \end{figure}

\begin{figure}
\caption{
 The crossover function for
 the staggered susceptibility, Eq.\ (\protect{\ref{chis_cross}}),
for various
values of $T_c / T_K $, normalized to its
value at $ T_K / 2 $.     }
  \label{fig_chis}
  \end{figure}

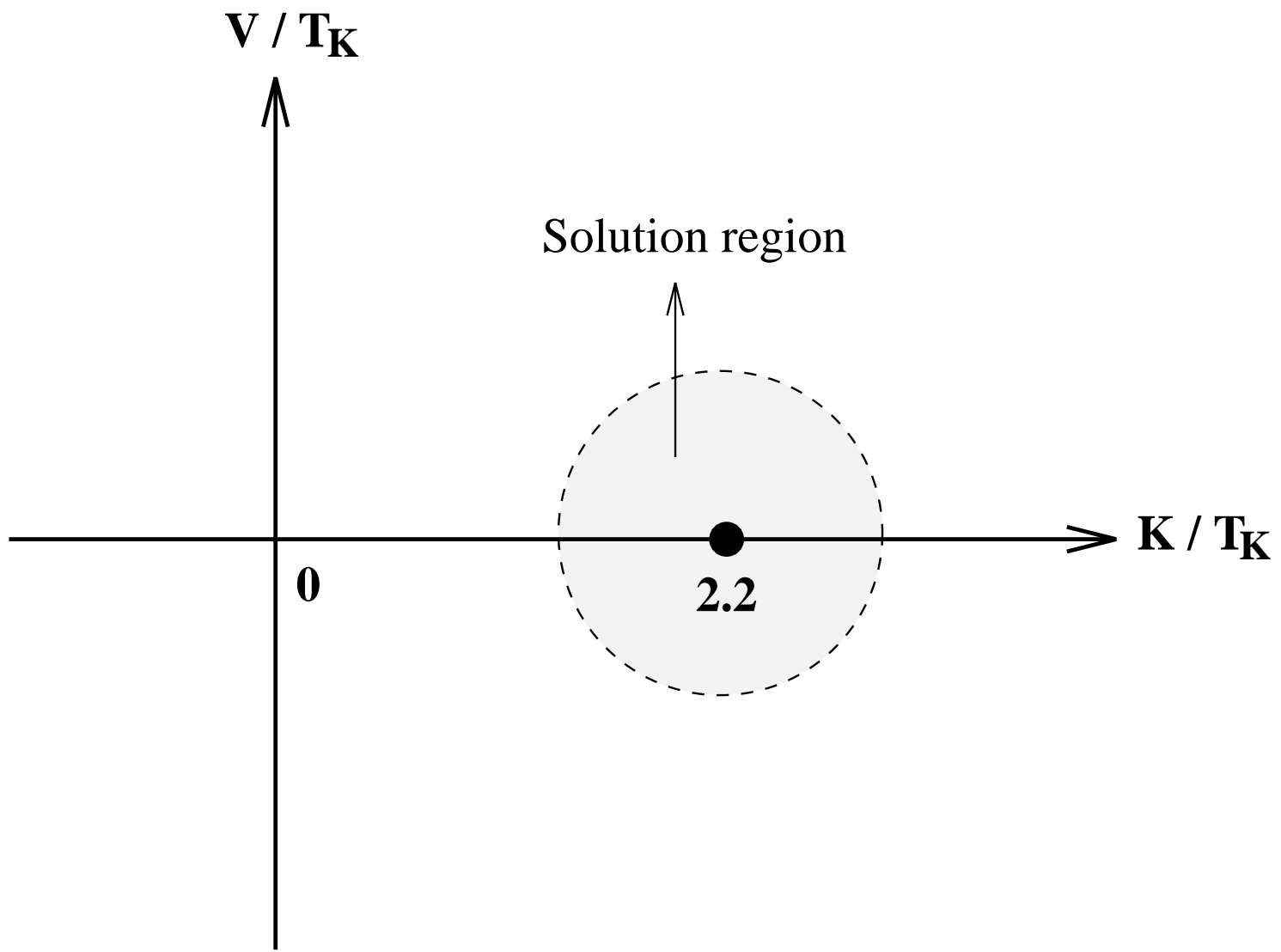
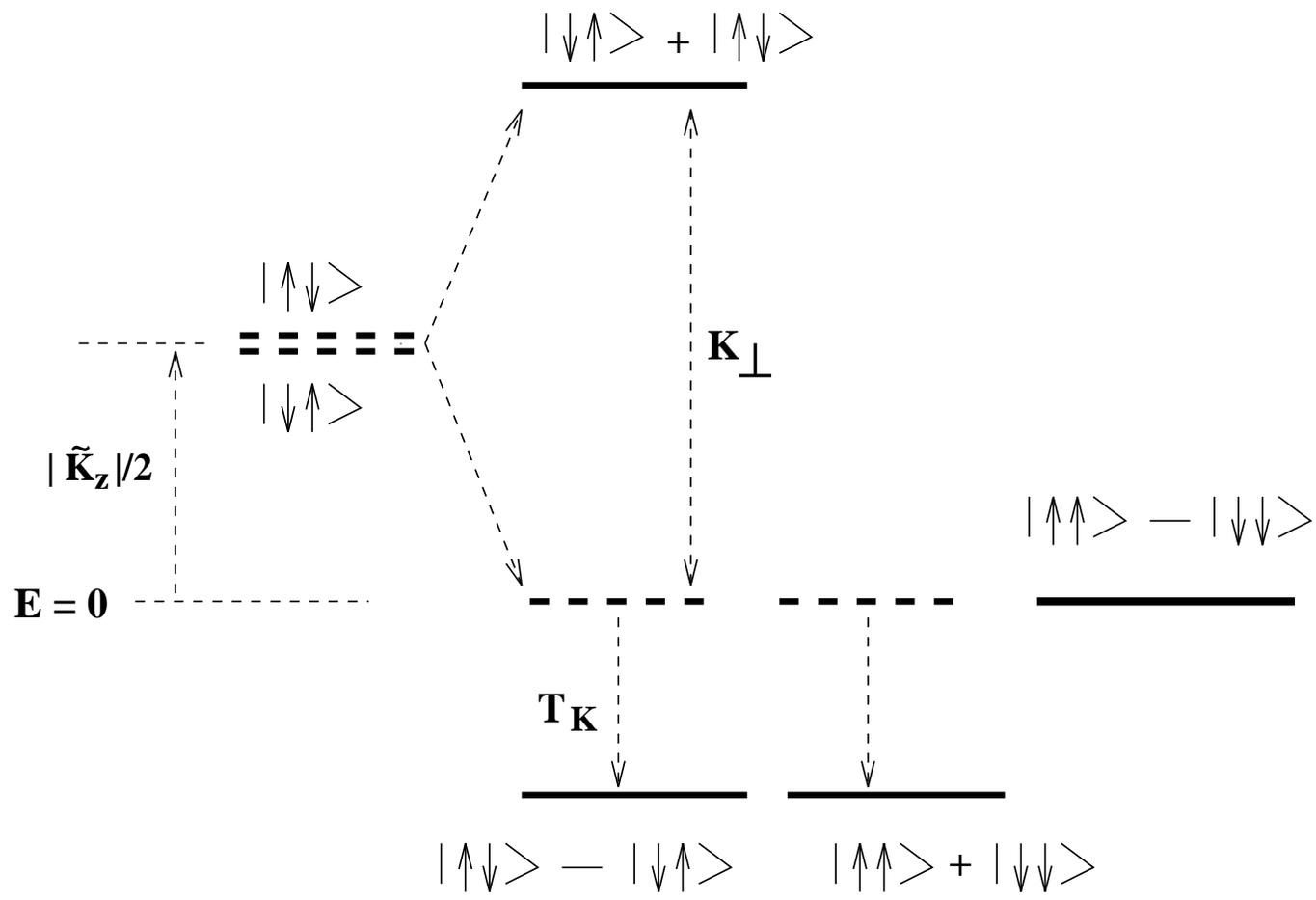
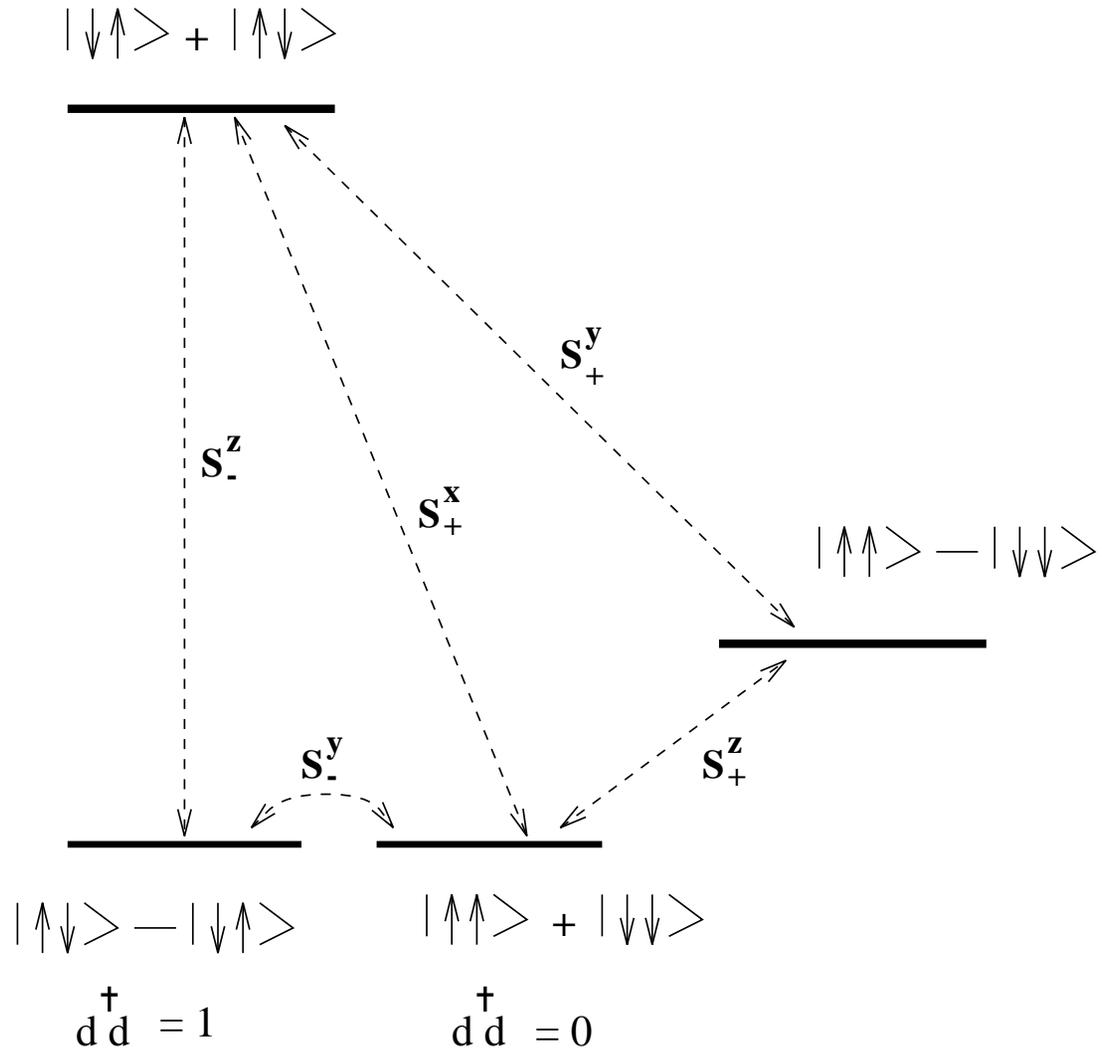
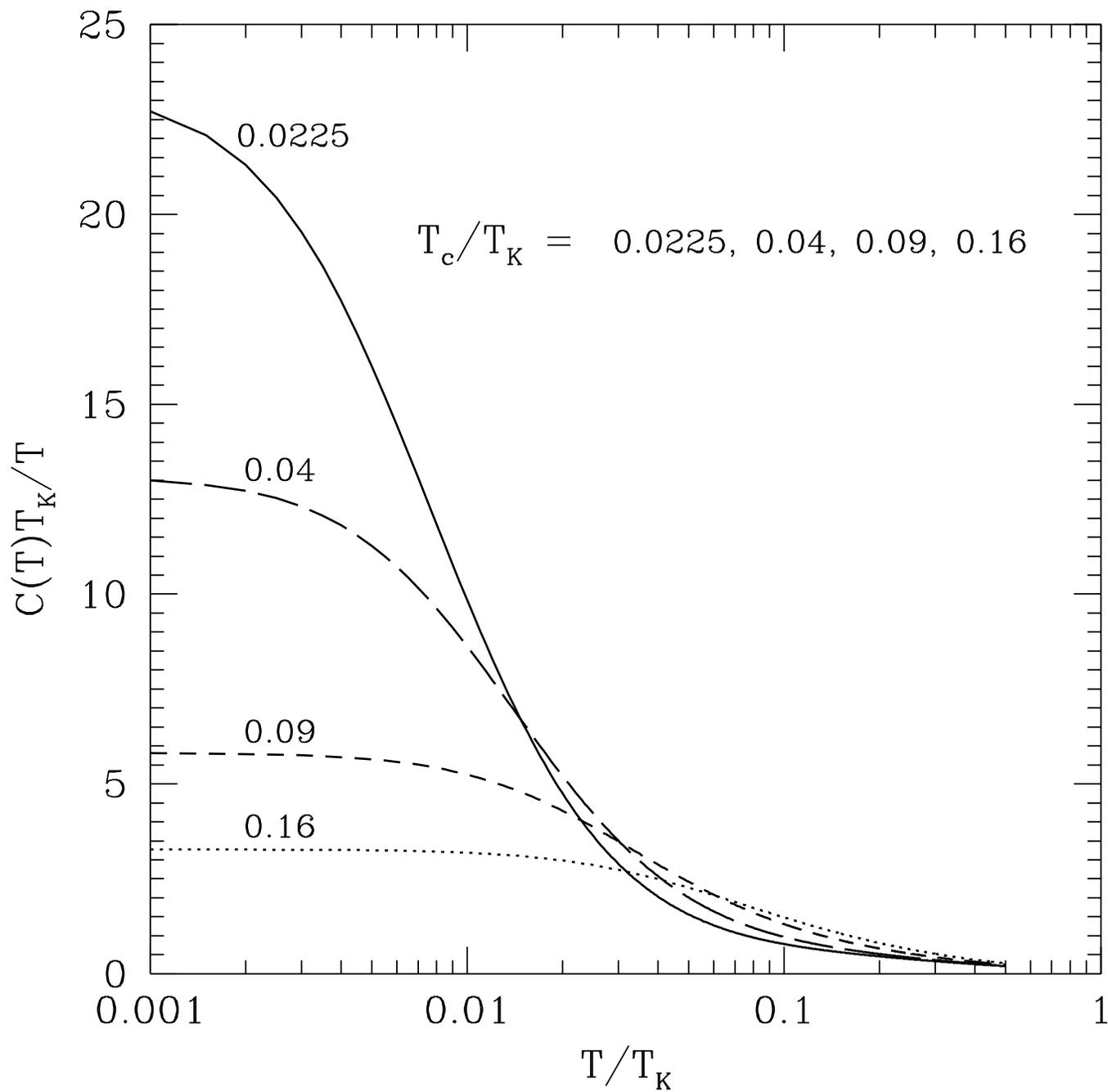
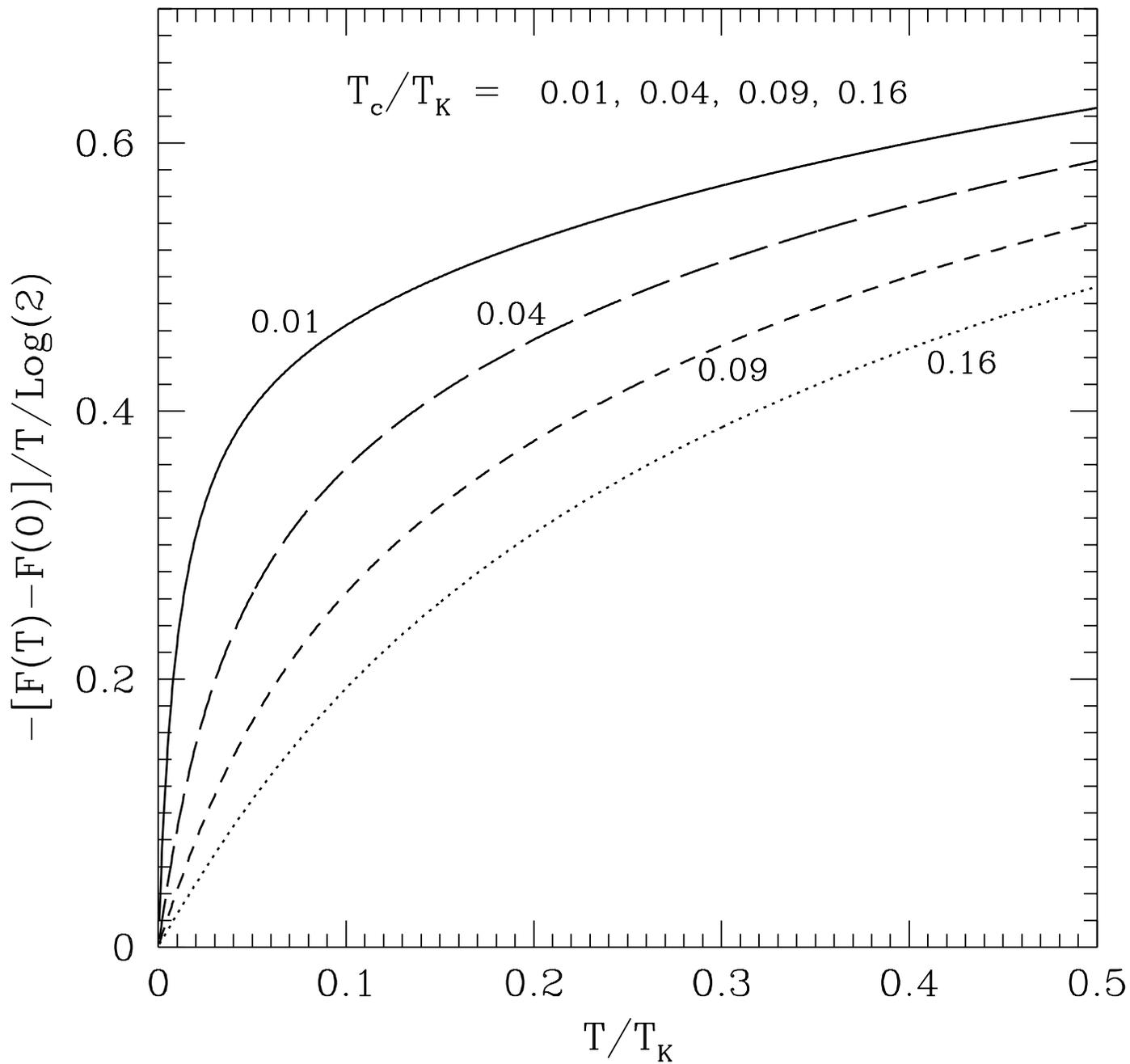
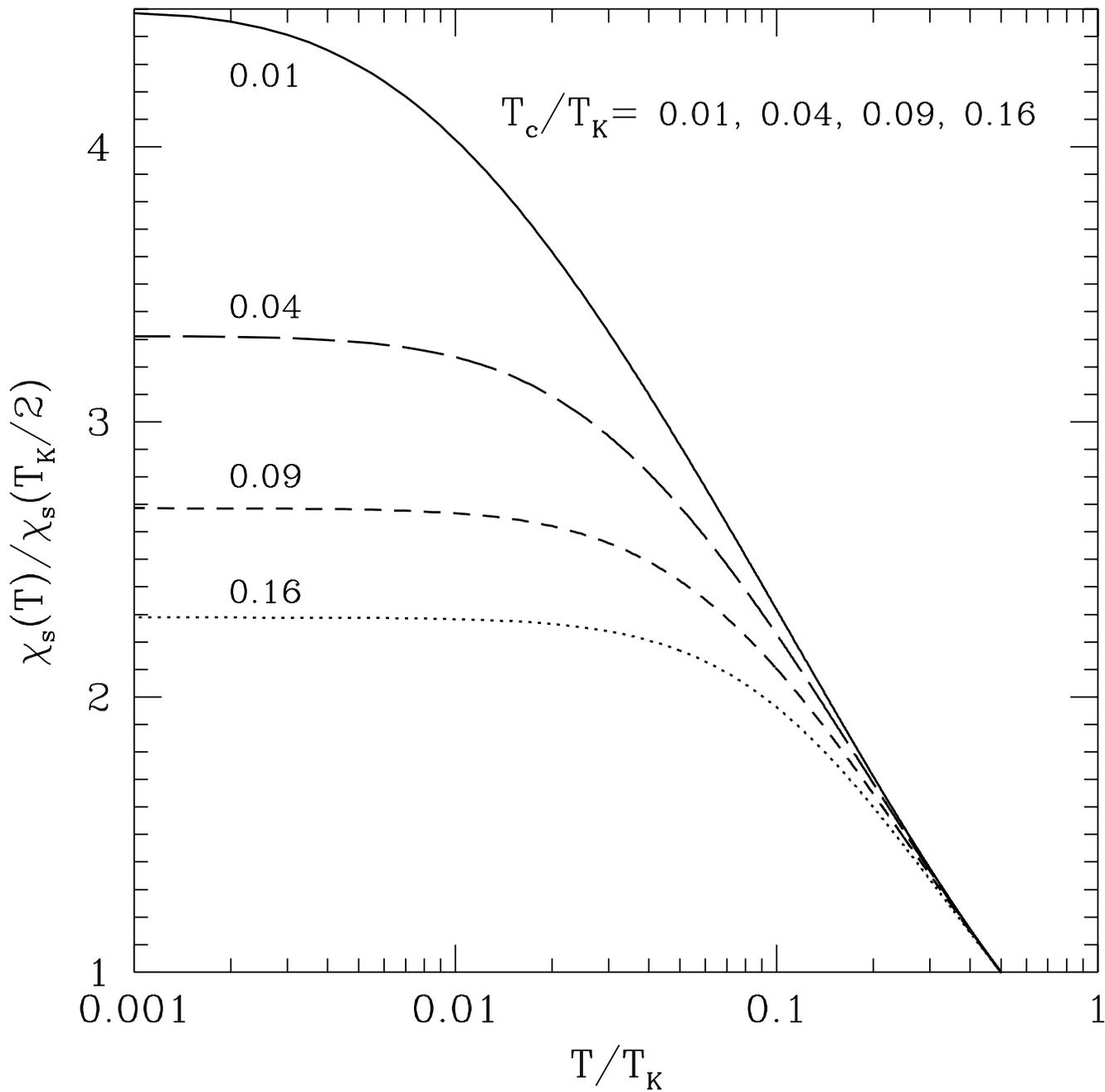
\end{document}